 \documentclass[preprint,12pt]{aastex}
 
\shorttitle{BD Kinematic Analysis}
\shortauthors{Faherty et al.}

\begin{document}


\title{The Brown Dwarf Kinematics Project (BDKP) I.  Proper Motions and Tangential Velocities for a Large Sample of Late-type M, L and T Dwarfs}


\author{Jacqueline K. Faherty\altaffilmark{1,3}, Adam J.\ Burgasser\altaffilmark{2}, Kelle L.\ Cruz\altaffilmark{4,6}, Michael M. Shara\altaffilmark{1}, Frederick M. Walter\altaffilmark{3}, and Christopher R. Gelino\altaffilmark{5}}

\altaffiltext{1}{Department of Astrophysics, 
American Museum of Natural History, Central Park West at 79th Street, New York, NY 10034; jfaherty@amnh.org }
\altaffiltext{2}{Massachusetts Institute of Technology, Kavli Institute for Astrophysics and Space Research,
Building 37, Room 664B, 77 Massachusetts Avenue, Cambridge, MA 02139}
\altaffiltext{3}{Department of Physics and Astronomy, Stony Brook University Stony Brook, NY 11794-3800}
\altaffiltext{4}{Astronomy Department, California Institute of Technology, Pasadena, CA 91125}
\altaffiltext{5}{Spitzer Science Center, California Institute of Technology, Pasadena, CA 91125}
\altaffiltext{6}{Spitzer Postdoctoral Fellow}



\begin{abstract}
We report  proper motion measurements for 427 late-type M, L and T dwarfs,
332 of which have been measured for the first time.  Combining these new proper motions
with previously published measurements yields  a sample of 841
M7-T8 dwarfs. We combined parallax measurements or calculated spectrophotometric distances and computed tangential
velocities for the entire sample.  We find that kinematics for the full and volume-limited 20 pc samples are consistent with those
expected for the Galactic thin disk, with no significant differences
between late-type M, L, and T dwarfs.  Applying an age-velocity relation we conclude that the average kinematic age of the 20 pc sample of ultracool dwarfs is older than recent kinematic estimates and more consistent with age results calculated with population synthesis models.  There is a statistically distinct population of high tangential velocity sources ($V_{tan}$  $>$ 100 km~s$^{-1}$)  whose kinematics suggest an even older population of ultracool dwarfs belonging to either the Galactic thick
disk or halo.  We isolate subsets of the entire sample, including low surface-gravity dwarfs, unusually blue L dwarfs, and photometric outliers in $J-K_{s}$ color and investigate their kinematics.  We
find that the spectroscopically distinct class of unusually blue L dwarfs has kinematics clearly consistent with old age, implying that high surface-gravity and/or low metallicity may be relevant to their spectral properties.  The low surface-gravity dwarfs are kinematically younger than the overall population, and the kinematics of the red and blue ultracool dwarfs suggest ages that are younger and older than the full sample, respectively.   We also present a reduced proper motion diagram at 2MASS $K_{s}$ for the entire population and find that a limit of  $H_{K_{s}}$ $>$ 18 excludes M dwarfs from the L and T dwarf population regardless of near-infrared color, potentially enabling the identification of the coldest brown dwarfs in the absence of color information.

\end{abstract}





\section{INTRODUCTION}

	 Kinematic analyses of stars have played a fundamental role in shaping our picture of the Galaxy and its evolution.  From early investigations (e.g. \citealt{1908Nachr} , \citealt{1925ApJ....62..191L}, and \citealt{1927BAN.....4...79O}) where the large scale structure of the Galactic disk was first explored, through more recent investigations (e.g. \citealt{1983MNRAS.202.1025G}, \citealt{1989ARA&A..27..555G},  \citealt{1998MNRAS.298..387D}, \citealt{2005A&A...430..165F}) where the structure of the Galaxy was refined to include a thick disk and prominent features such as streams, moving groups, and superclusters, kinematics have played a vital role in understanding the Galactic origin, evolution, and structure.  Combining kinematics with spectral features, several groups have mapped out ages and metallicities for nearby F,G,K, and M stars (e.g. \citealt{2004A&A...418..989N}).  The ages of these stars have become an important constraint on the Galactic star formation history and their kinematics have become a vital probe for investigating membership in the young thin disk, intermediate aged thick disk or older halo portion of the Galaxy.    
	\\ \indent   One population that has yet to have its kinematics exploited is the recently discovered population of very low-mass ultracool dwarfs (UCDs).  These objects, which include those that do not support stable hydrogen fusion (\citealt{1962AJ.....67S.579K}; \citealt{1963PThPh..30..460H}), occupy the late-type M through T dwarf spectral classifications  \citep[e.g.,][and references therein]{2005ARA&A..43..195K}.    UCDs emit the majority of their light in the infrared and thus were only discovered in large numbers with the advent of wide-field near-infrared imaging surveys such as the Two Micron All Sky Survey (hereafter 2MASS; \citealt{2006AJ....131.1163S}), the Deep Infrared Survey of the Southern Sky (hereafter DENIS; \citealt{1997Msngr..87...27E}) and the Sloan Digital Sky Survey (hereafter SDSS; \citealt{2000AJ....120.1579Y}).  Their very recent discovery has largely precluded astrometric measurements which require several-year baselines to produce useful measurements. Therefore, while UCDs appear to be comparable in number to stars (e.g., \citealt{1999ApJ...521..613R}), their role in the structure of the Galaxy has yet to be explored.  
	\\ \indent   In addition, the thermal evolution of brown dwarfs (the lowest temperature ultracool dwarfs) implies that there is no direct correlation between spectral type and mass, leading to a mass/age degeneracy which makes it difficult to study the mass function and formation history of these objects.  While some benchmark sources (e.g. cluster members, physical companions to bright stars) have independent age determinations, and spectroscopic analyses are beginning to enable individual mass and age constraints (e.g. \citealt{2006ApJ...639.1095B}; \citealt{2007ApJ...656.1136S}, \citealt{2004ApJ...609..885M}), the majority of brown dwarfs are not sufficiently characterized to break this degeneracy.  Kinematics can be used as an alternate estimator for the age of the brown dwarf population.
	\\ \indent Moreover,  kinematics can also be used to characterize subsets of UCDs.  With hundreds of UCDs now known\footnote{An up-to-date list of known L and T dwarfs is maintained by C. Gelino, D. Kirkpatrick and A. Burgasser at http://www.dwarfarchives.org}, groupings of peculiar objects \-- sources whose photometric or spectroscopic properties differ consistently from the majority of the population \-- are becoming distinguishable.  Currently defined subgroups of late-type M, L and T dwarfs include 1) low surface-gravity, very low-mass objects (e.g. \citealt{2004ApJ...600.1020M}, \citealt{2006ApJ...639.1120K}, \citealt{2007ApJ...657..511A}, \citealt{2007AJ....133..439C}), 2) old, metal-poor ultracool subdwarfs (e.g. \citealt{2003ApJ...592.1186B}, \citealt{2003ApJ...591L..49L}, \citealt{2006AJ....132.2372G}, \citealt{2007ApJ...658..557B}), 3) unusually blue L dwarfs (hereafter UBLs; e.g. \citealt{2003AJ....126.2421C}, \citealt{2007AJ....133..439C}; \citealt{2004AJ....127.3553K}, \citealt{2006AJ....131.2722C}), and 4) unusually red and possibly dusty L dwarfs (e.g. \citealt{2008arXiv0806.1059L}; \citealt{2003ApJ...596..561M}).  While observational peculiarities can overlap between these groups (e.g. both young and dusty L dwarfs can be unusually red), they appear to encompass objects with distinct physical traits (e.g., mass, age, composition, and cloud properties) so they are important for drawing a connection between observational characteristics and intrinsic physical properties.  Kinematics can be used to investigate the underlying physical causes for the peculiarities of these groups.  
		\\ \indent 	In the past decade a number of groups have conducted astrometric surveys of UCDs, including subsets of low mass objects  (e.g.  \citealt{2004AJ....127.2948V}, \citealt{2002AJ....124.1170D}, \citealt{2000AJ....120.1085G}, \citealt{2003AJ....126..975T}, \citealt{2007AJ....133.2258S}, \citealt{2007arXiv0710.4786J}, \citealt{2007ApJ...666.1205Z}, and \citealt{2008AJ....135..785W,2006AJ....132.2507W}). We have initiated the Brown Dwarf Kinematics  Project (BDKP) which aims to measure the positions and 3D velocities of all known L and T dwarfs within 20 pc of the Sun and selected sources of scientific interest at larger distances (e.g. low surface-gravity dwarfs, subdwarfs).  In this article we add 332 new proper motion measurements and combine all published proper motion measurements and distance estimates into a uniform sample to examine the ultracool dwarf population as a whole.  Section 2 of this paper outlines the observed sample and describes how proper motion measurements were made.  Section 3 discusses the expanded sample and how distances and $V_{tan}$ measurements were calculated.   Section 4 examines the full astrometric sample and subsets. Section 5 reviews the high tangential velocity objects in detail.  Finally, section 6 applies an age-velocity relation and examines resultant ages of the full sample and red/blue outliers.

\section{OBSERVATIONS AND PROPER MOTION MEASUREMENTS}
\subsection{$Sample$ $Selection$}
\begin{figure*}[htbp]
\begin{centering}
\includegraphics[width=1.0\hsize]{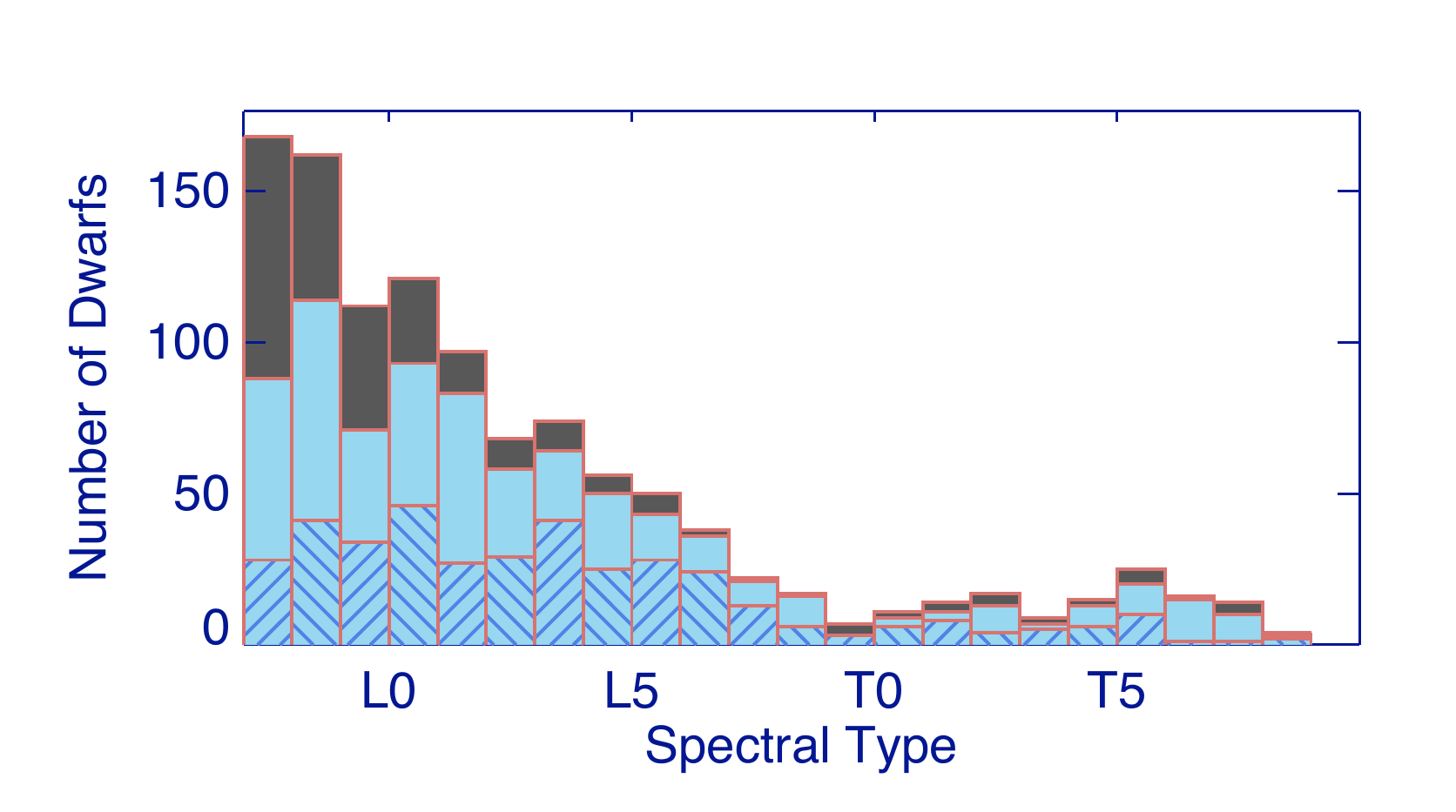}
\end{centering}
\caption{ Spectral type distribution of all late-type M, L, and T dwarfs.  The overall histogram is the distribution of all ultracool dwarfs in our sample.  The blue shaded histogram shows ultracool dwarfs with proper motion measurements.  The diagonally shaded histogram shows the distribution of ultracool dwarfs with new proper motions reported in this paper.
 } 
\label{fig:SpT_Dist}
\end{figure*}

	Our goal is to re-image all known late-type M, L, and T dwarfs to obtain accurate uniformly measured proper motions for the entire ultracool dwarf population.  In our sample we focused on the lowest temperature L and T dwarfs that were lacking proper motion measurements or whose proper motion uncertainty was larger than 40 mas/yr.  We gave high priority to any dwarf that was identified as a low surface-gravity object in the literature.  Our sample was created from 634 L and T dwarfs listed on the Dwarf Archives website as well as 456 M7-M9.5 dwarfs gathered from the literature (primarily from Cruz et al. 2003, 2007).  The sample stayed current with the Dwarf Archives website through April 2008.  Figure ~\ref{fig:SpT_Dist} shows the histogram of spectral type distributions for the entire sample.  The late-type M and early-type L dwarfs clearly dominate the ultracool dwarf population.  Plotted in that figure is the current distribution of objects with proper motion values and the distribution of objects for which we report new proper motions.   To date we have re-imaged 427 objects.   As of June 2008 and including all of the measurements reported in this article, 570 of the 634 known L and T dwarfs and 277 of the 456 late-type M dwarfs in our sample have measured proper motions.

\subsection{$Data$ $Acquisition$ $and$ $Reduction$ }

Images for our program were obtained using three different instruments and telescopes in the northern and southern hemispheres.  Table 1 lists the instrument properties.  For the northern targets the 1.3m telescope at MDM with the TIFKAM IR imager in J band was used. For the southern targets the 0.9m and 1.5m telescopes at CTIO with the CFIM optical imager in I band and the CPAPIR wide field-IR imager in J band (respectively) were used.  The CTIO data were acquired through queue observing   on 11 nights in March, September and December 2007, and standard user observing on 9 nights in January 2008.  The MDM targets were imaged on five nights in November 2007 and 7 nights in April 2008.   Objects were observed as close to the meridian as possible up to an airmass of $1.80$, and with seeing no greater than 2.5$''$ FWHM.   Exposure times varied depending on the target and the instrument.  For CPAPIR the exposure times ranged over 15$\--$40s with 4 co-adds per image and a five-point dither pattern.  At MDM the exposure times ranged over 30$\--$120s with up to 6 co-adds per image and a three to five-point dither pattern.  For the 0.9m observations the exposure times ranged over 180$\--$1800s per image with no co-adds and a three-point dither pattern.  The dither offset between positions with each instrument was 10$''$.  

All data were processed in a similar manner using standard IRAF and IDL routines.  Domeflats were constructed in the J or I band. CPAPIR and CFIM domeflats were created from 10 images illuminated by  dome lamps, and TIFKAM domeflats were created by subtracting the median of 10 images taken with all dome lights off from the median of 10 images taken with the dome lights on.  A dark image constructed from 25 images taken with the shutter closed was used to map the bad pixels on the detector.  Domeflats were then dark subtracted and normalized.  Sky frames were created for each instrument by median-combining all of the science data taken on a given night.  Science frames were first flat-fielded, then sky-subtracted. Individual frames were shifted and stacked to form the final combined images.  

\subsection{$Calculating$ $Proper$ $Motions$}

The reduced science frames were astrometrically calibrated using the 2MASS Point Source catalogue.  2MASS astrometry is tied to the TYCHO 2 positions and the reported astrometric accuracy varies from source to source.  In general the positions of 2MASS sources in the magnitude range $9<K_{s}<14$ are repeatable to 40-50 mas in both RA and DEC.  

Initial astrometry was fit by inputting a 2x2 transformation matrix containing astrometry parameters which were first  calculated from one image in which two stars with known 2MASS RA and DEC values and second epoch X,Y pixel positions were known. The reference RA, DEC and pixel values were first set to the pointing RA and DEC values and the center of the chip respectively.  

RA and DEC values for all stars in the field were then imported from the 2MASS point source catalogue and converted to X,Y pixel positions using the initial astrometric parameters. We worked in X,Y positions of the second epoch image so we could overplot point source positions on an image and visually check that we converged upon a best fit solution.  We detected point sources on the second epoch image with a centroiding routine which used a detection threshold of 5 sigma above the background.  We matched the 2MASS X,Y positions to the second epoch positions by cross-correlating the two lists. We refined the astrometric solution by a basic six parameter, least-squares, linear transformation where we took the positions from the 2MASS image ($X_{1}$,$Y_{1}$) and the positions from the second epoch image ($X_{0}$,$Y_{0}$) and solved for the new X,Y pixel positions of the second epoch image in the 2MASS frame. Due to the large field of view, we checked for higher order terms in the CPAPIR images and found no significant terms.  The following equations were used:   
\begin{equation}
        X = x_{2o} + A(X_1-X_0)+B(Y_1-Y_0)
\end{equation}
\begin{equation}
Y=y_{2o}+C(X_1-X_0)+D(Y_1-Y_0)
\end{equation}
where $x_{2o}$ and $y_{2o}$ were set to the center of the field, $A,B,C,D$ solve for the rotation and plate scale in the two coordinates. 

The sample of stars used to compute the astrometric solution for each image were selected according to the following criteria:  
\begin{enumerate}

\item Only stars in the 2MASS J magnitude range $12<J<15$ were used, as objects in this intermediate magnitude range transformed with the smallest residuals from epoch to epoch.

\item The solution reference stars were required to transform with total absolute residuals against 2MASS of $< 0.2$ pixels.  From testing with images taken consecutively using each instrument, the best astrometric solution always generated between 0.1 and 0.2 pixel average residuals.  Therefore the stars used to calculate the solution were required to fall in or below that range.

\end{enumerate}

As the solution was iterated, the residuals were examined at each step, and stars that did not fit the above criteria were removed.  For CPAPIR, the process converged on a solution that had between 100 and 200 reference stars with average residuals $< 0.15$ pixels.    TIFKAM and CFIM have smaller fields of view ($\sim$6 arcmin and $\sim$ 5 arcmin respectively as opposed to 35 arcmin for CPAPIR) so there were far fewer stars to work with.  For these imagers the process converged on a solution that had between 15 and 60 reference stars.  The astrometric solution was required to converge with no less than 15 reference stars and when this criterion could not be met, the other two criteria above were relaxed.  As a result TIFKAM and CFIM had slightly larger residuals on the astrometric solution (average residuals $< 0.25$ pixels).
 
Once an astrometric solution was calculated, final second epoch positions were computed using a Gaussian fit for each 2MASS X,Y position on an image.  For the science target, a visual check was employed to ensure that it had been detected and X,Y positions were manually input for the Gaussian fit.  Final X,Y positions were then converted back into RA and DEC values using the  best astrometric solution and the proper motion was calculated using the positional offset and time difference between the second epoch image and 2MASS.  

\begin{figure*}[!ht]
\begin{center}
\epsscale{1.5}
\plottwo{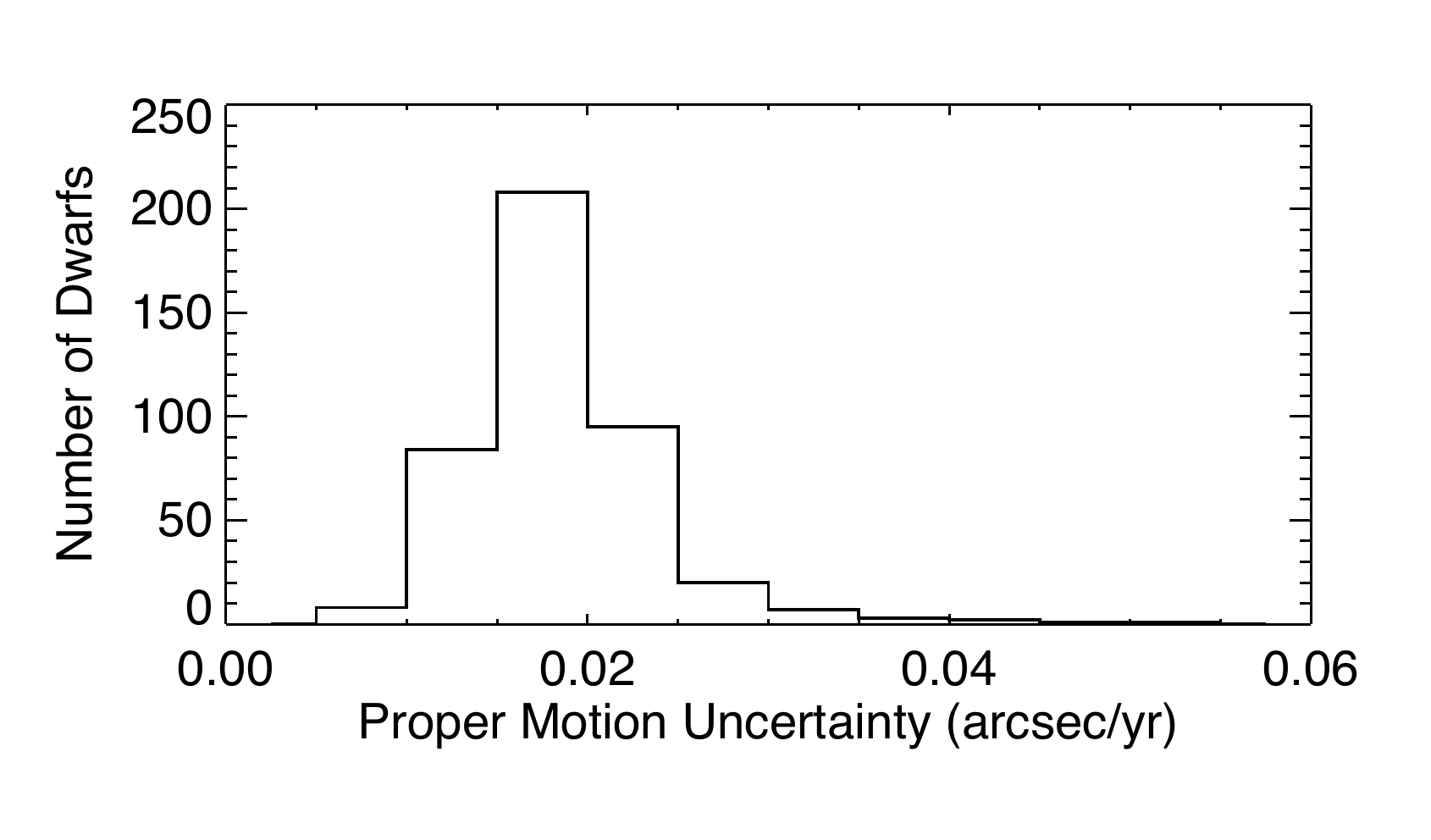}{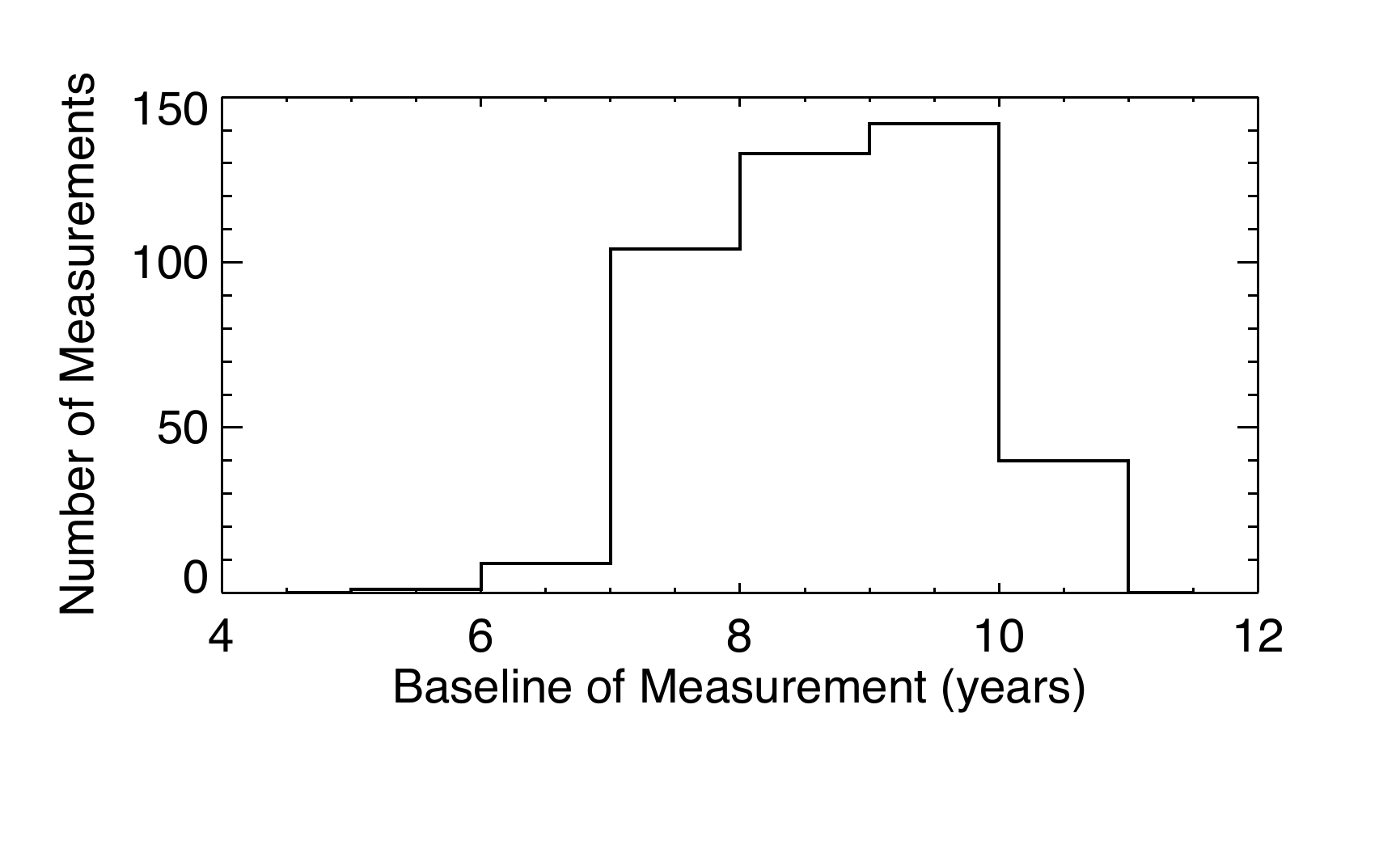}
\end{center}

\caption{ (Top): Distribution of proper motion uncertainties for the sample of 427 measurements reported in this paper.  The median value is 18 mas/yr . (Bottom): Distribution of proper motion baselines (time between first and second epoch measurements) used in this survey.} 
\label{fig:Uncertainty}
\end{figure*}

The residuals of the astrometric solution were converted into proper motion uncertainties by multiplying by the plate scale of the instrument and then dividing by the epoch difference.  The baselines ranged from 6-10 years and our astrometric uncertainties range from 5 to 50 mas/yr.  Positional uncertainties for each source were also calculated by comparing the residuals of transforming the X,Y position for our target over consecutive dithered images.  These uncertainties are dominated by counting statistics with the high S/N sources having negligible positional uncertainties compared to the uncertainties in the astrometric solution.  We added the positional and astrometric solution uncertainties in quadrature to determine the total proper motion uncertainty.   Figure~\ref{fig:Uncertainty} shows the distribution of proper motion uncertainties and baselines for all new proper motion measurements reported in this paper.  The median uncertainty was 18 mas/yr. 

\begin{figure*}[!ht]
\begin{center}
\epsscale{1.5}
\plottwo{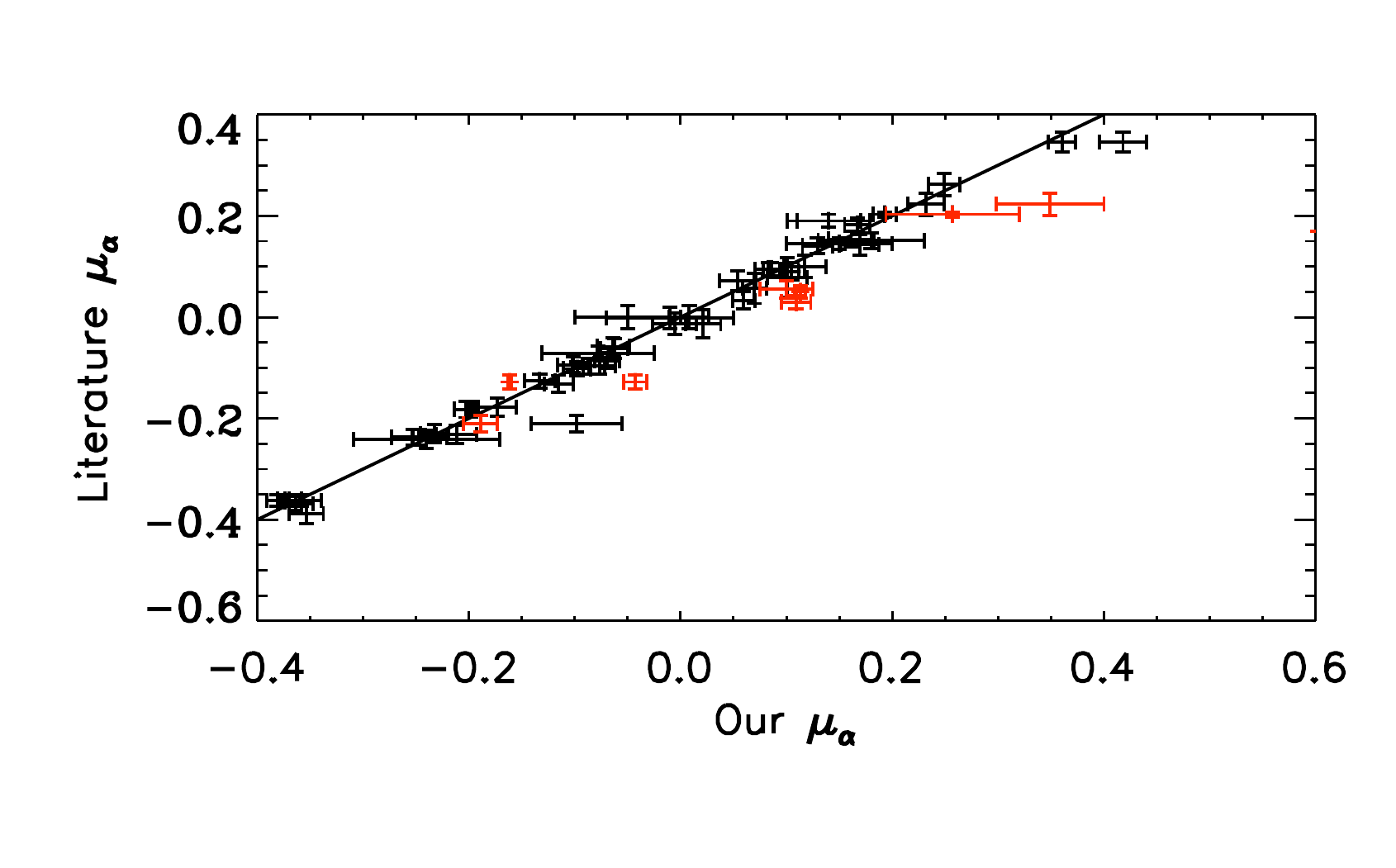}{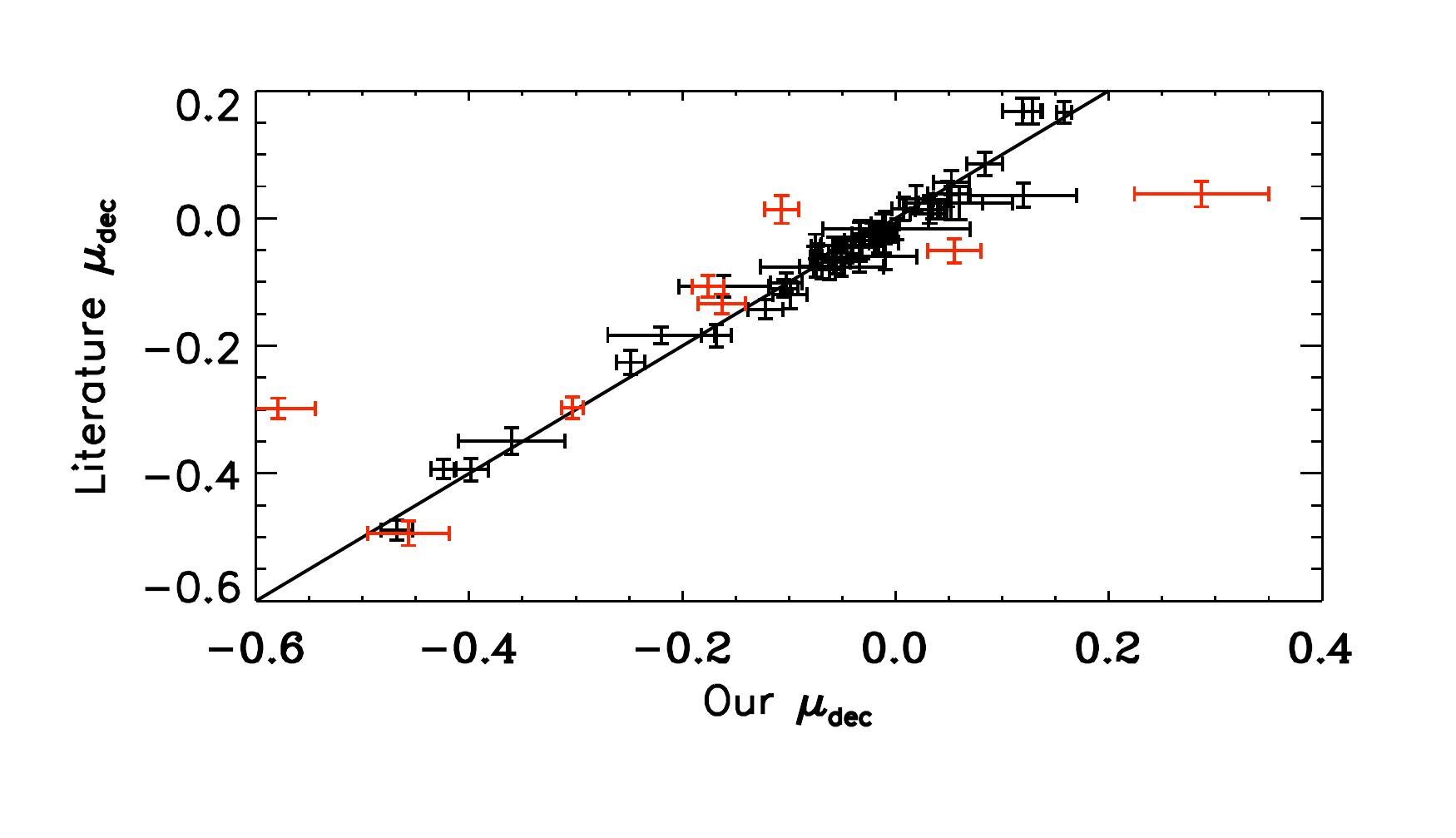}
\end{center}

\caption{ Comparison of Right Ascension (top) and Declination (bottom) proper motion measured in this paper and those measured in the literature.  The straight line represents a perfect agreement between measurements.  The red highlighted objects are the discrepant proper motion measurements (see  Table 4.)} 
\label{fig:Mu_Check}
\end{figure*}

Of the 427 proper motion measurements we report in this paper,  332 are presented here for the first time.  Twelve objects were purposely remeasured with multiple instruments as a double check on the accuracy of the astrometric solution,  and 42 objects were remeasured to refine the proper motion uncertainties. Thirty-two measurements were published in \citet{2007arXiv0710.4786J}\--hereafter J07\--, and 11 in \citet{2007ApJ...667..520C} while our observations were underway.  The proper motion measurements presented in this paper agree  to better than $2\sigma$ in  84 of the 97 cases of objects with prior measurements.  Table 4 lists those cases where the proper motions are discrepant by more than $2\sigma$ with a published value.  For nine of the objects, there is a third (fourth or fifth) measurement by an independent group with which we are in good agreement.   We are discrepant with six objects reported in \citet{2005A&A...435..363D} but we note that there are no position angle uncertainties reported for these objects in that catalogue therefore we cannot fully assess the accuracy of the proper motion components.  The difference in proper motion for 2MASSW J1555157-095605 is quite large ($>$ 1''/yr difference) but there are two other measurements for this object with which we are in close agreement.   We have examined all of the discrepant proper motion images carefully and see no artifacts that could have skewed our measurements.  Figure ~\ref{fig:Mu_Check}  compares the proper motion component measurements from this paper with those from the literature for objects with $\mu$ $<$ 0.5 $\arcsec$/yr and $\mu_{err}$ $<$  0.1 $\arcsec$/yr.  With$\sim$90\% agreement with published results, this indicates that the 332 new measurements are robust. Table 2. contains all new measurements reported in this article, and Table 3. contains the astrometric measurements for the full sample.

\section{DISTANCES, TANGENTIAL VELOCITIES, AND REDUCED PROPER MOTION}

\subsection{$Expanded$ $Sample$}
We extended our observational sample to include published late-type M, L, and T dwarfs with proper motion measurements yielding a  full combined sample containing 841 objects.  Thirty-three percent of ultracool dwarfs in the full sample have multiple proper motion measurements.   In these cases we chose the measurement with the smallest uncertainty for our kinematic analysis, typically objects from high precision astrometric surveys such as Vrba et al 2004 or  Dahn et al 2002.  If there was a value discrepant by more than 2$\sigma$ amongst multiple measurements ($>$ 2) for an object then regardless of uncertainty we defaulted to the numbers that were in agreement and chose the one with the smaller uncertainty.   Otherwise, if there was a discrepancy and only two measurements, we quoted the one that had the smaller uncertainty and made note of it during the analysis.  

\subsection{$Distances$ $and$ $Tangential$ $Velocities$}
\indent  True space velocities are a more fundamental measure of an object's kinematics than apparent angular motions, so proper motions for the complete sample were converted to tangential velocities using astrometric or spectrophotometric distances.  As of January 2008, only 79 of the 634 L and T dwarfs and 64 of the 456 late-type M dwarfs in our sample had published parallax measurements.  Therefore to include the other 83\% of L and T dwarfs and 87\% of late-type M dwarfs in a population analysis, published absolute magnitude/spectral type relations were used for calibrating distances.   \citet{2002AJ....124.1170D} and \citet{2004AJ....127.2948V} both showed  that $M_J$ is well correlated with spectral type for late-type M, L, and T dwarfs (see also\citealt{2005PASP..117..706W}, and \citealt{2007AJ....134.2398C}). Since the initial relations were published several investigators have revised the absolute magnitude/spectral type relation after including new measurements and removing resolved binaries.  In this paper, the distances for  the M7-L4.5 dwarfs were calculated using  the absolute 2MASS J magnitude/spectral type relation in \citet{2003AJ....126.2421C} and the distances for  the L5 --T8  dwarfs were calculated using  the absolute MKO K magnitude/spectral type relation in \citet{2007ApJ...659..655B}\footnote{The coefficients of this polynomial relation reported in
\citet{2007ApJ...659..655B} did not list sufficient significant digits, yielding a
slightly different numerical relation than that used in the paper's
analysis.  The coefficients as defined should be \{$c_i$\} = [10.4458,
0.232154, 0.0512942, -0.0402365, 0.0141398, -0.00227108, 0.000180674,
-6.98501e-06, 1.05119e-07], where $M_{K}$ = $\sum_{i=0}^{6}{c_i}$SpT$^i$
and SpT(T0) = 10, SpT(T5) = 15, etc.}.  Both optical and near-IR  spectral types are reported for ultracool dwarfs.  For late-type M through the L dwarfs, we use the optical spectral type in the distance relation when available but use near-IR spectral types when no optical spectral types are reported. We use the near-IR spectral type in the distance relation for all T dwarfs.  The Cruz et al. (2003) relation was derived for the 2MASS magnitude system, while the  \citet{2007ApJ...659..655B} relation was derived using the MKO system.  In reporting distances we maintain the magnitude system for which the relation was calculated, converting a 2MASS magnitude to an MKO magnitude or vice versa using the relation in \citet{2004PASP..116....9S}  when necessary.  The most recent precision photometry for many L and T dwarfs (e.g. Knapp et al 2004; Chiu et al. 2006, 2008) are reported on the MKO system; yet the majority of objects explored in this paper have measured 2MASS magnitudes.  We convert MKO filter measurements to the 2MASS system when available using the conversion relations of \citet{2004PASP..116....9S} so all of the ultracool dwarf photometry in Table 3 is reported on the 2MASS system.
\\ \indent 	The uncertainty in the derived distance is dominated by the uncertainty in the spectral type (the photometric uncertainties are typically between 0.02\--0.1 mag whereas the spectral type uncertainties are typically 0.5\--1.0).  This leads to a systematic over- or underestimation of distance of up to 30\%.  Therefore the kinematic results presented in this paper are largely sensitive to the reliability of the spectrophotometric distances used to calculate $V_{tan}$.  Furthermore, unresolved multiplicity leads to an underestimation of distance.  Recent work has shown that roughly 20\% of ultracool dwarfs are likely to be binary (\citealt{2007ApJ...668..492A}, \citealt{2008AJ....135..580R}), and this fraction may be even higher across the L dwarf/T dwarf transition (\citealt{2006ApJS..166..585B}).  Seven percent (56) of the dwarfs analyzed in this paper are known to be close binaries and of these, most appear to be near equal-mass/equal brightness (e.g. \citealt{2003AJ....126.1526B}; \citealt{2006ApJS..166..585B}).  For these objects we use the distances quoted in the binary discovery papers where the contribution of flux from the secondary was included in the distance estimate. Any remaining tight binaries probably constitute no more than 10-20\% of the sample and the contamination of their inclusion in the kinematic analysis is relatively small.

 \subsection{$Reduced$ $Proper$ $Motion$ $Diagram$}
A reduced proper motion diagram is a useful tool for distinguishing between kinematically-distinct stellar and substellar populations.   This parameter was used extensively in early high proper motion catalogues to explore Galactic structure (\citealt{1973IAUS...54...11L}).  Proper motion is used as a proxy for distance measurements following the expectation that objects with large proper motions will be nearest to the Sun.  The definition is analogous to that of absolute magnitude:

\begin{equation}
H=m+5.0+5.0 log_{10}(\mu)
\end{equation}
or
\begin{equation}
H=M+5.0 log_{10}(V_{tan})-3.38
\end{equation}
where m and M are the apparent and absolute magnitudes (respectively), $V_{tan}$ is measured in km/s and $\mu$ is measured in $\arcsec$/yr.

\begin{figure*}[htbp]
\begin{centering}
\includegraphics[width=1.0\hsize]{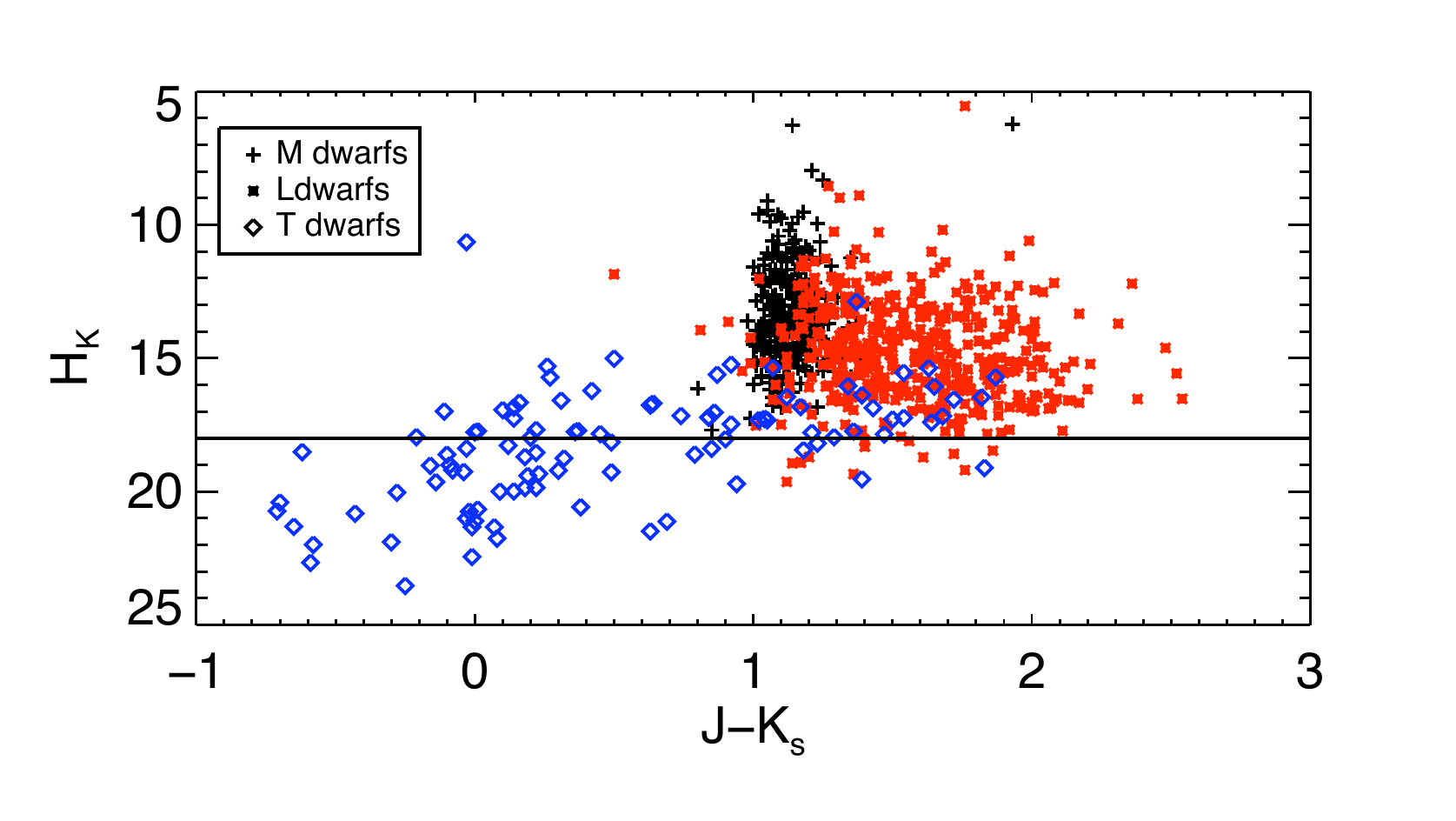}
\end{centering}
\caption{ Reduced proper motion diagram using the 2MASS J and $K_{s}$ magnitudes.  Late-type M dwarfs are marked with a black plus sign, L dwarfs are marked as a red 5 point star and T dwarfs are marked as blue diamonds.  The line at $H_{K}$ of 18 marks where M dwarfs are segregated from the L and T dwarfs regardless of near-IR color.  This cut-off will also include subdwarfs and cool white dwarfs but these objects will be rare.} 
\label{fig:PM_REDUC}
\end{figure*}

We can use reduced proper motion to search for the lowest temperature objects.  In Figure~\ref{fig:PM_REDUC}, we show the reduced proper motion at $K_{s}$ for our astrometric sample.  We find that below an $H_{K_{s}}$ of 18 there are only L and T dwarfs regardless of near-IR color.  Since the discovery of the first brown dwarfs, near-IR color selection has been the primary technique for identifying strong candidates.  But because M dwarfs dominate photometric surveys (they are bright, nearby and found in large numbers),  near-IR color cut-offs were administered to maximize the L and T dwarfs found in searches.  These cut-offs have caused a gap in the near-IR color distribution of the brown dwarf population, particularly around $J -K_{s}$ of 1 where early-type T dwarfs and metal weak L dwarfs are eliminated along with M dwarfs.  A reduced proper motion diagram with the cut-off limit cited above allows a search which eliminates the abundant M dwarfs and probes the entire range of $J-K_{s}$ colors for the ultracool dwarf population.

Note that, while our cut-off limits are good guidelines for segregating the coolest temperature dwarfs within the ultracool dwarf population, there is likely to be contamination in selected regions of the sky from relatively rare ultracool subdwarfs and cool white dwarfs, which are nonetheless of scientific interest.  


\section{ANALYSIS}

\subsection{$Kinematic$ $Characteristics$ $of$ $the$ $Ultracool$ $Dwarf$ $Population$}
The ultracool dwarfs analyzed in this paper have a range of proper motion values from 0.01$\--$4.7$''$/yr and a range of proper motion uncertainties from 0.0002 $\--$ 0.3 $''$/yr.   While one of our goals is to refine proper motion measurements of ultracool dwarfs to have uncertainties $<$ 40 mas/yr, there are still 86 or 10\% that have larger errors.  Since the uncertainty in $V_{tan}$ is generally dominated by the uncertainty in distance (see subsection 3.2 above) we make no restrictions on the accuracy of the proper motion measurements used in the kinematic analysis.  The median $1\sigma$ detection limit for proper motion measurements in this paper was 18 mas/yr (see Figure ~\ref{fig:Uncertainty}).  We use this number as a proxy for the L and T dwarfs (where we are looking at all known field objects as opposed to the late-type M dwarfs where we are looking at only a subset) to determine the percentage of objects with appreciable motion. We find that 32 move slower than our 2$\sigma$ detection limit and ten of those are at or below our 1$\sigma$ limit. This indicates that according to our astrometric standard, less than 6\% of L and T dwarfs have no appreciable motion.  Conversely, 32 objects (or 6\% of the population) move faster than 1.0$''$/yr making them some of the fastest proper motion objects known. As late-type dwarfs are intrinsically quite faint and have only been detected at nearby distances (generally $\leq$ 60~pc), the high proper motion values measured are not surprising.  Using the median proper motion values listed in Table 5 as a proxy , we can also conclude that at least half or more of the brown dwarf population would be easily detectable on a near-IR equivalent of Luyten's 2-tenth catalog (\citealt{1979lccs.book.....L}) where the limiting proper motion was $\sim$ 0.15 $\arcsec$/yr.  

Table 5 lists the average proper motion values and photometric data for the entire population binned by spectral type.  There is a trend within these data for larger proper motion values with increasing spectral type.  This is clearest within the L0-L9 population where the sample is largest.  We further bin this group into thirds to compare a statistically significant sample.  We examine the L0-L2, L3-L5, and L6-L9 populations and find the median proper motion values increase as  (0.174, 0.223, 0.289) $\arcsec$/yr.  This trend most likely reflects the fact that earlier type sources are detected to further distances.  Indeed when we examine the median distance values for these same groupings we find values of (31, 27, 20)~pc respectively.

 \subsection{$Kinematics$ $of$ $Full$ $and$ $20pc$ $Samples$}
 
\begin{figure*}[htbp]
\begin{centering}
\includegraphics[width=1.0\hsize]{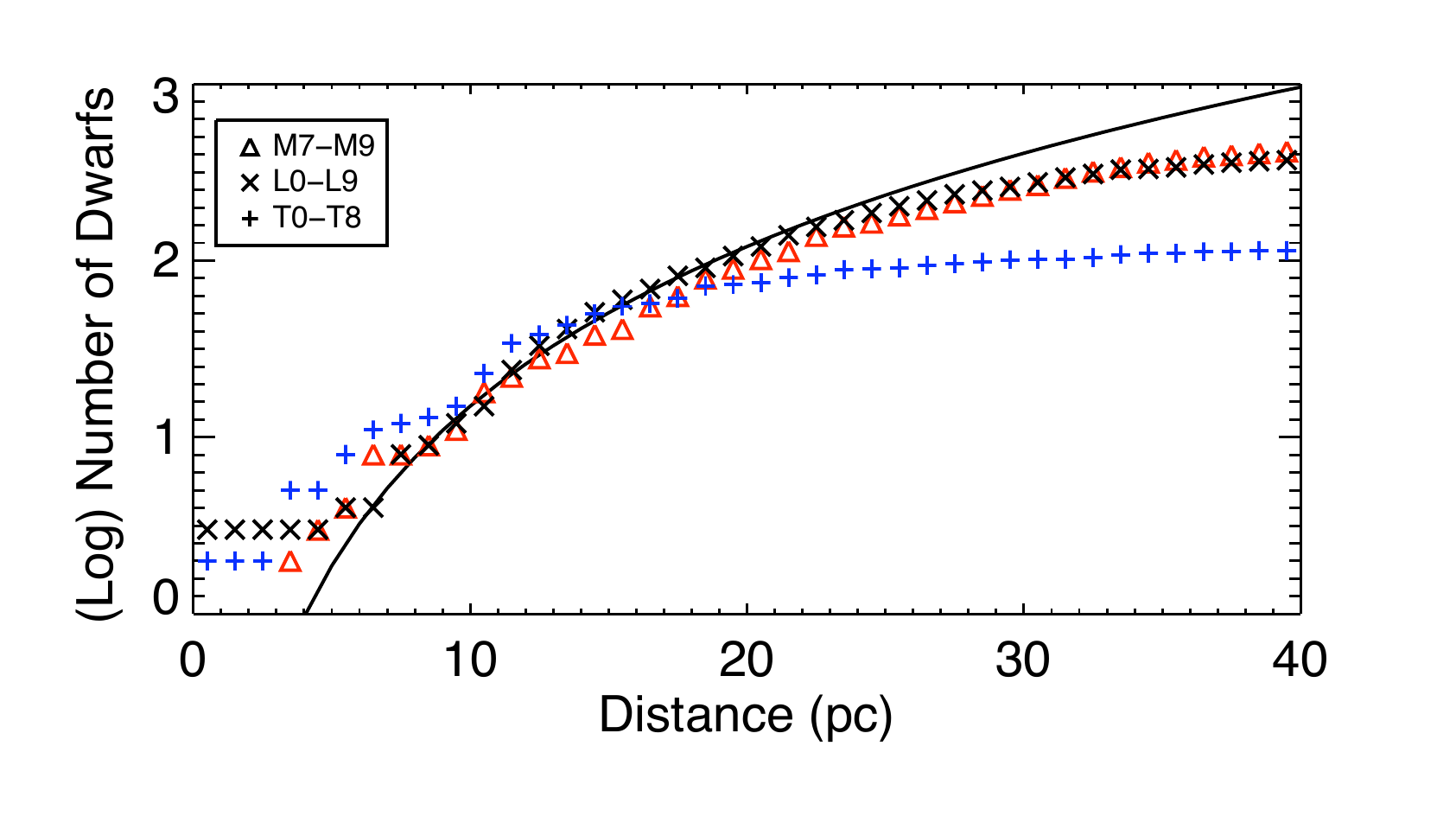}
\end{centering}
\caption{ Cumulative distance distribution of all late-type M, L, and T dwarfs in our database.  Triangles refer to the M7-M9 dwarfs, the 'X' symbols refer to all L0-L9 dwarfs and the plus symbols refer to all T0-T8 dwarfs.  The solid line corresponds to a constant density distribution (N $\propto$ d$^3$).  The  L and M dwarfs deviate from this distribution around 20 pc but the T dwarfs fall off closer to 15 pc. } 
\label{fig:Dist_Distr}
\end{figure*}

  We have conducted our kinematic analysis on two samples: the full astrometric sample and the 20 pc sample. Figure ~\ref{fig:Dist_Distr} shows the distance distribution for all ultracool dwarfs regardless of proper motion measurements to demonstrate the pertinence of the 20 pc sample.  In this figure, both the late-type M and L dwarfs diverge from an  N $\propto$ d$^3$ density distribution around 20 pc. The T dwarfs diverge closer to 15 pc.   Within the literature (e.g Cruz et al. 2003) complete samples to 20 pc have been reported through mid-type L dwarfs so we use this distance in order to establish a volume-limited kinematic sample.  We also examine the two samples with and without objects with $V_{tan}$ $>$ 100 km~s$^{-1}$ in order to remove extreme outliers that may comprise a different population.  

\begin{figure}[!ht]
\begin{center}
\epsscale{1.5}
\plottwo{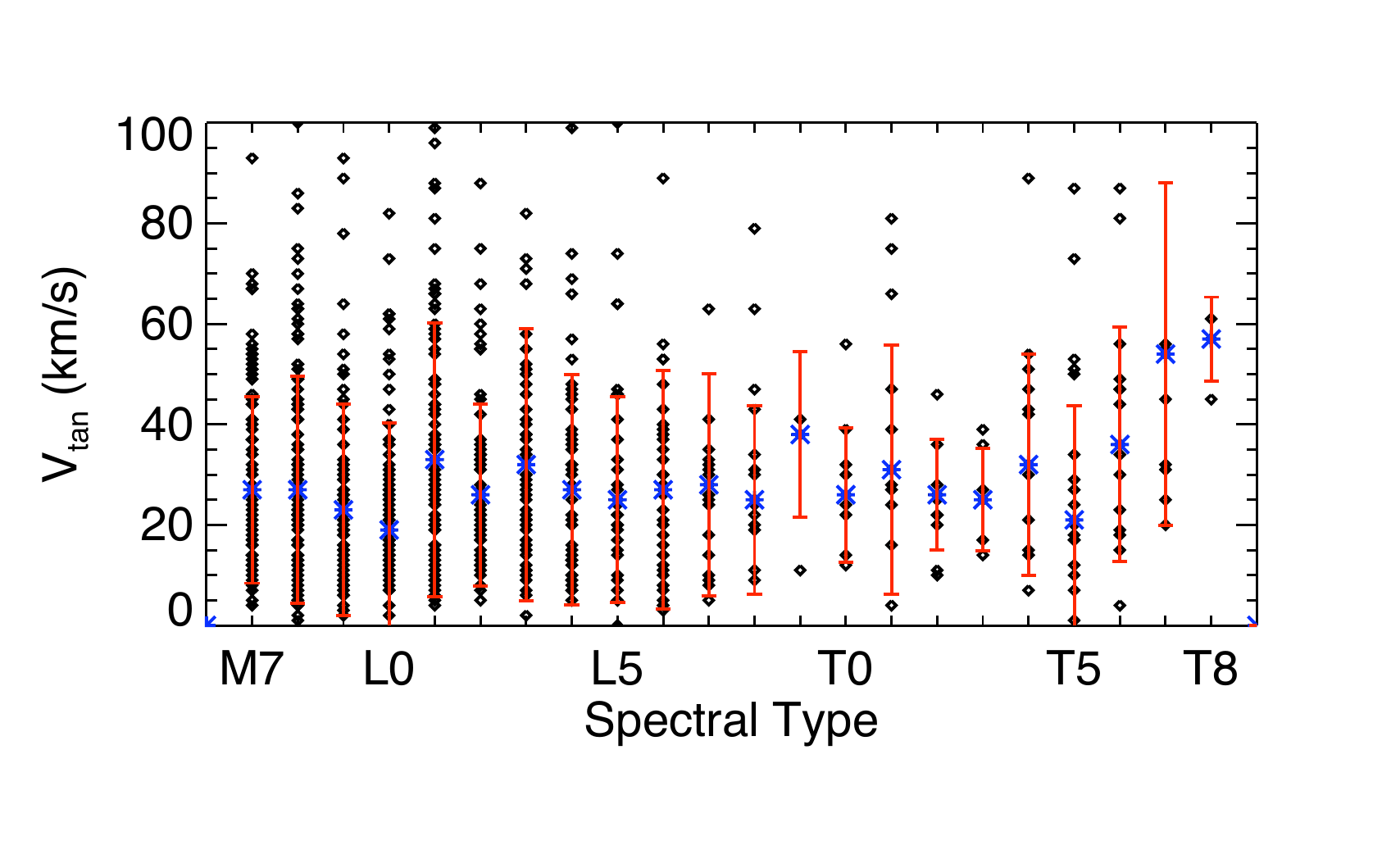}{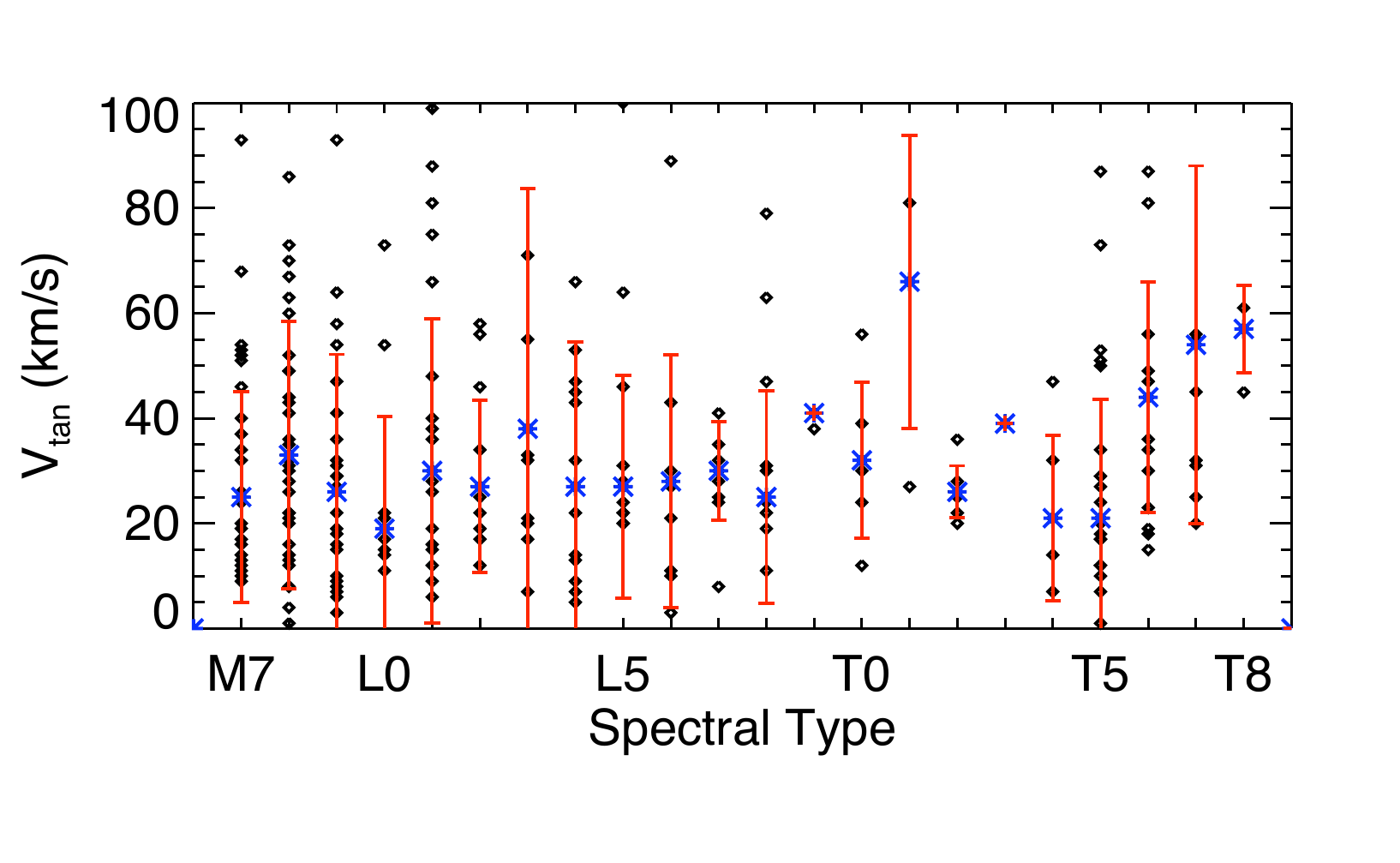}
\end{center}
\caption{ Distribution of $V_{tan}$ values binned by spectral type. The top panel is the full astrometric sample and the bottom panel is the 20 pc sample.  The asterisks refer to the median $V_{tan}$ values and the vertical bars refer to the standard deviation or dispersion of velocities.
 } 
 \label{fig:TAN_DIS}
\end{figure}

Tables  6 \-- 7 contain the mean kinematic properties for the 20 pc sample and the full astrometric sample respectively. Figure ~\ref{fig:TAN_DIS} shows $V_{tan}$ vs. spectral type for  both samples.  As demonstrated in Figure ~\ref{fig:TAN_DIS}, we find no difference between the two samples, with median $V_{tan}$ values of 26 km~s$^{-1}$  and  29 km~s$^{-1}$ and $\sigma_{tan}$ values of 23 km~s$^{-1}$ and 25 km~s$^{-1}$ for M7-T9 within the full sample and the 20 pc sample respectively.   Within spectral class bins, namely the M7-M9, L0-L9, or T0-T9 groupings, we find no significant kinematic differences. This indicates that we are sampling a single kinematic population regardless of distance and spectral type. 

\begin{figure*}[htbp]
\begin{centering}
\includegraphics[width=1.0\hsize]{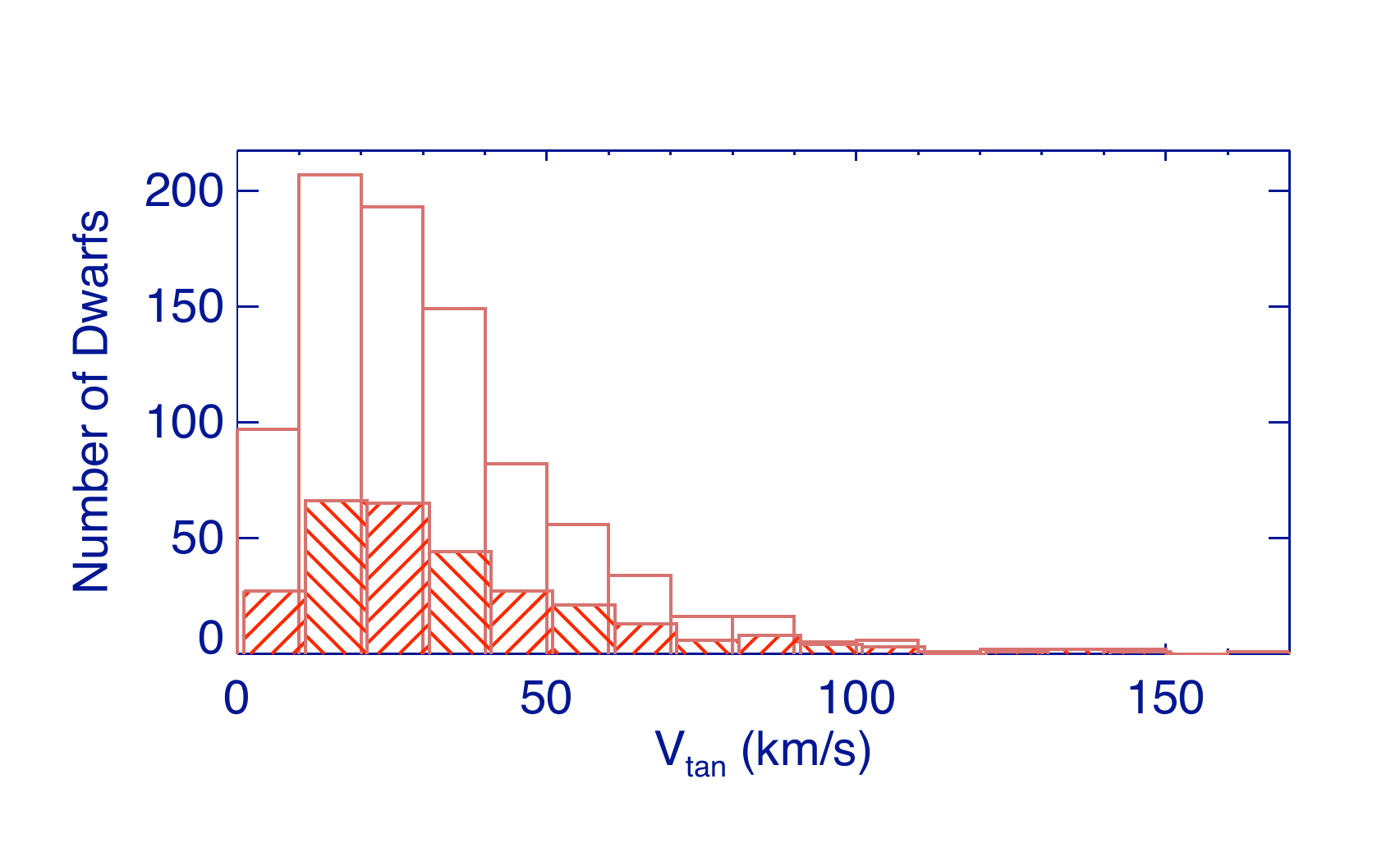}
\end{centering}
\caption{   The overall histogram is the tangential velocity distribution for the entire sample and the diagonally shaded histogram is the 20 pc sample.  Both $V_{tan}$ distributions peak in the 10-30 km~s$^{-1}$ bins.
 } 
\label{fig:vtanall}
\end{figure*}

Figure ~\ref{fig:vtanall} shows the distribution of tangential velocities.  There are 14 objects with $V_{tan}$ $> 100$ km~s$^{-1}$ that fall at the far end of the distribution.  Exclusion of these high velocity dwarfs naturally reduces the median $V_{tan}$ and $\sigma_{tan}$ values. The most significant difference in their exclusion occurs within the L0-L9 group as 10 of the 14 objects belong to that spectral class.   We explore the importance of this subset of the ultracool dwarf population in section~5.  

In order to put our kinematic measurements in the context of the Galaxy we compare with Galactic U,V,W dispersions.  Proper motion, distance, and radial velocity are all required to compute these space velocities.  Therefore, a direct Galactic U,V,W comparison with the ultracool dwarf population is not possible because radial velocity measurements for ultracool dwarfs are sparse, with only 48 of the L and T dwarfs to date having been reported in the literature (e.g. \citealt{2003ApJ...583..451M},  \citealt{2007ApJ...666.1205Z}, \citealt{2004A&A...419..703B}).  This is a similar problem to that for precise brown dwarf parallax measurements, but there is no relationship for  estimating radial velocities as there is for estimating distances.  However, we can divide our sample into three groups along Galactiocentric coordinate axes (toward poles, in direction of Galactic rotation and radially to/from Galactic center)  in order to minimize the importance of radial velocity in 2 out of the 3 space velocity components.  We create cones of 0 (all inclusive), 30, and 60 degrees around the galactic X,Y, and Z axes. Inside of each cone we set either the U,V, or W velocity to zero if the cone surrounds the galactic X,Y or Z axis respectively.  In this way we can set the radial velocity of each source to zero with minimum impact on the component velocities of the entire sample and gather U,V,W information for the known ultracool dwarf population.  We emphasize that this analysis is crude as the distribution of ultracool dwarfs is not isotropic (the Galactic plane has largely not been explored), and while the cones help to minimize the importance of radial velocity  unless an object is directly on the X,Y, or Z axis, the radial velocity component will contribute to the overall velocities.  Therefore, the spread of U,V,W velocities will be biased toward a tighter dispersion than the true values.  In order to calculate total velocities ($V_{tot}$) for objects, which requires U,V, and W velocities we choose a cone of 30 degrees which provides a statistically significant sample.  We create the cone around the X,Y, or Z axis and assume that within that cone either the V,W or U,Z or U,V components respectively are correct.  To obtain the third component we assume it to be the average of the two calculated ones.  In this way we can gather $V_{tot}$ information which will be used for age calculation purposes in Section 6.

\begin{figure}[!ht]
\begin{center}
\epsscale{1.2}
\includegraphics[width=.55\hsize]{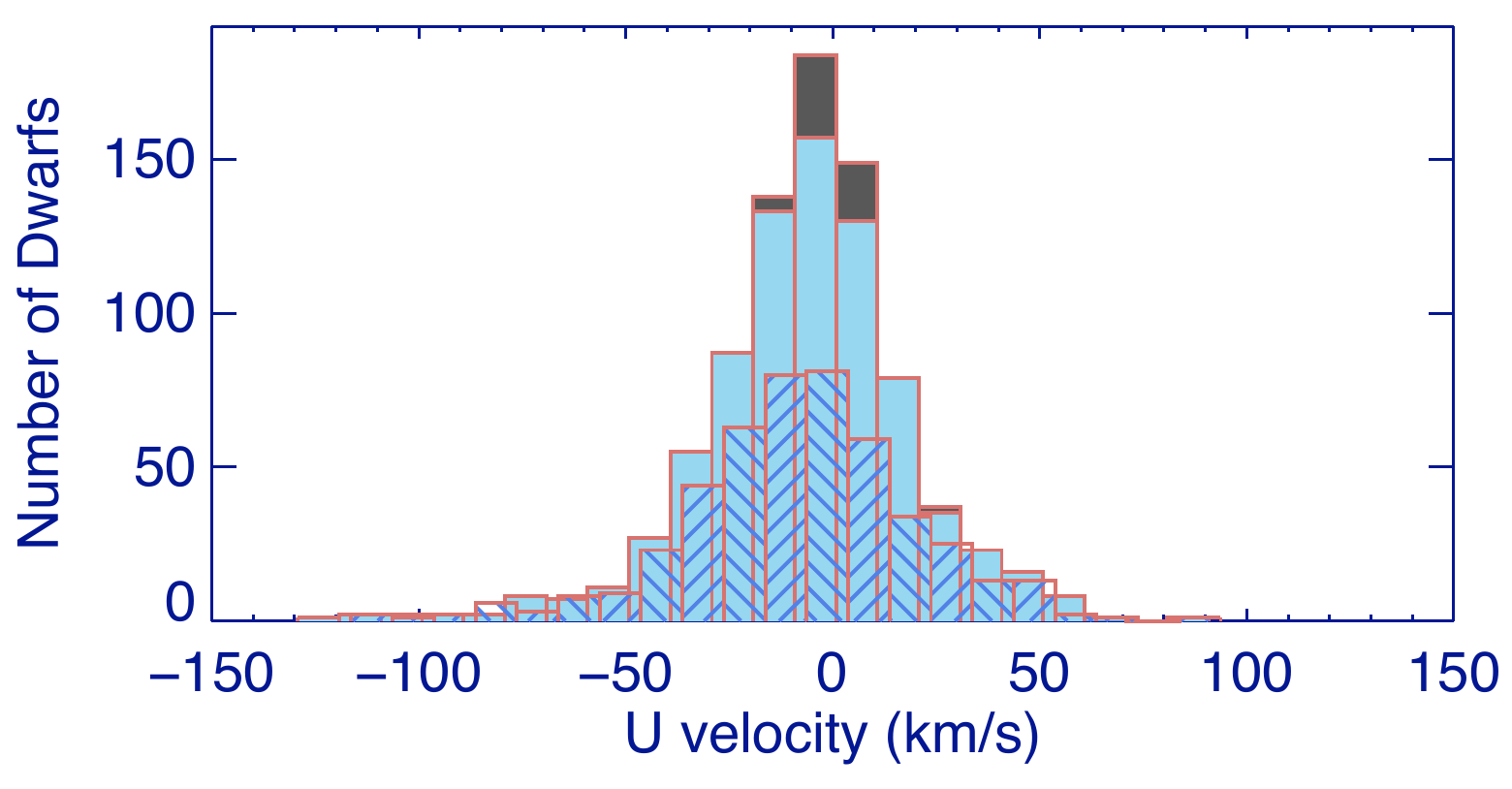}
\plottwo{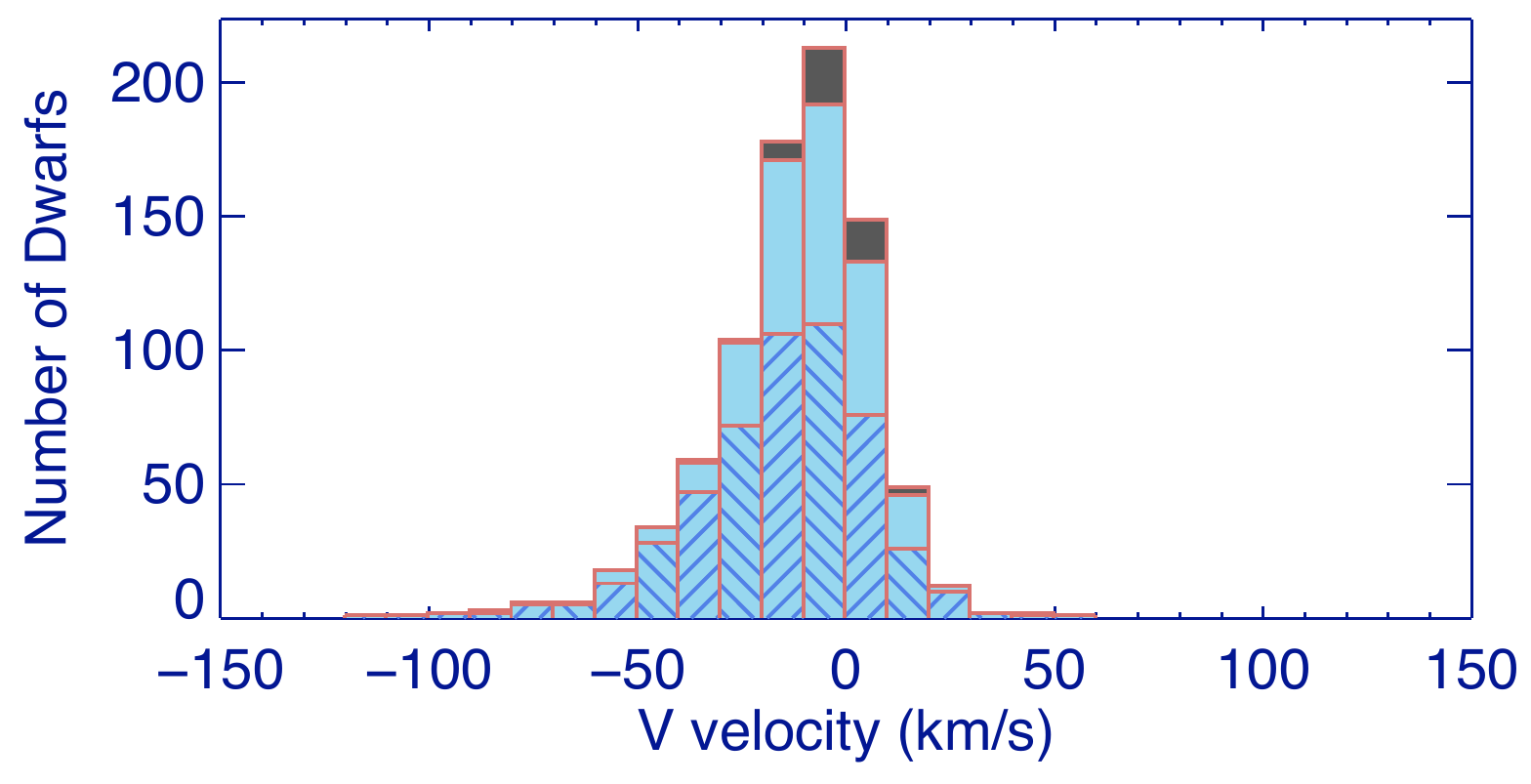}{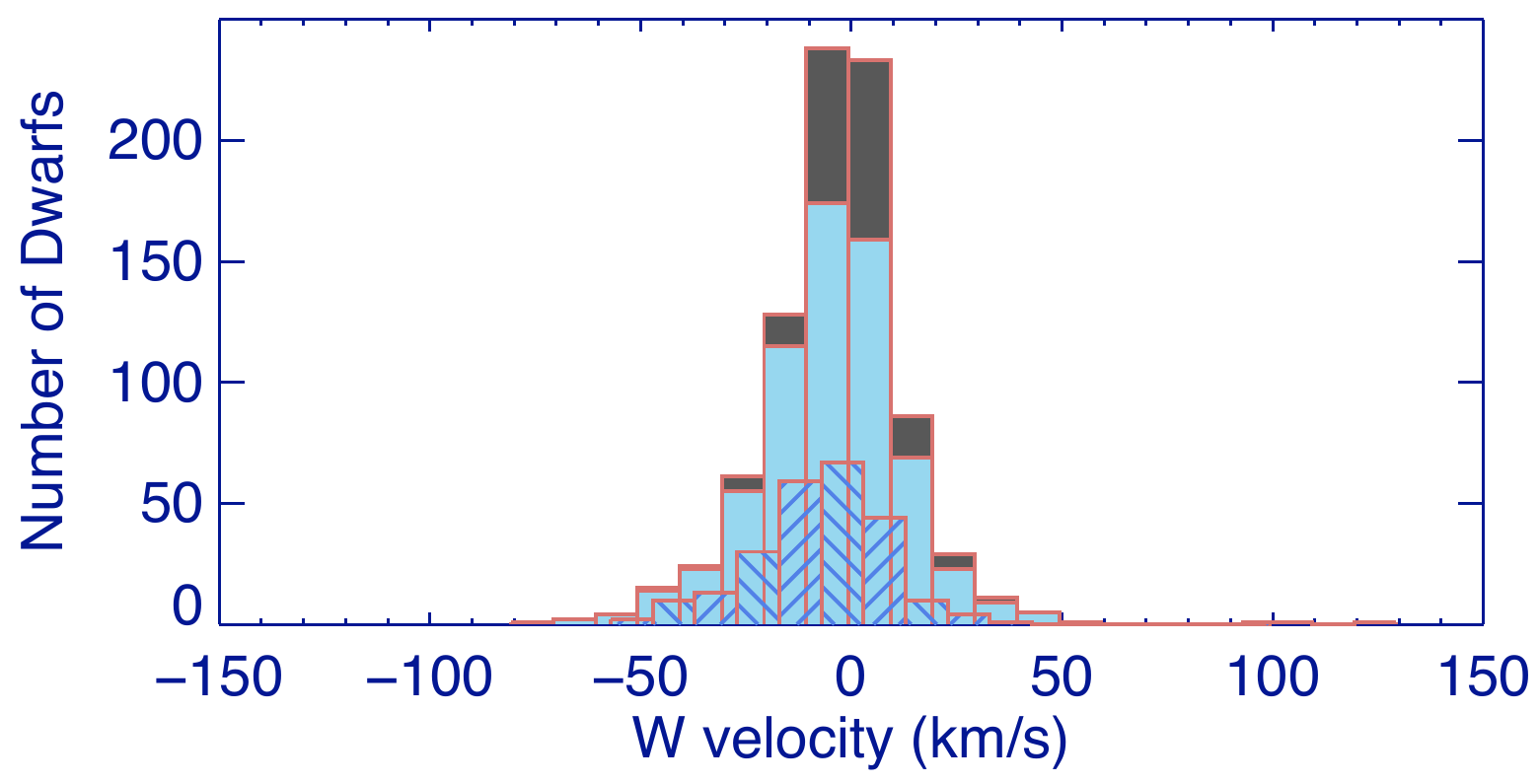}
\end{center}
\caption{ Histogram of U,V,W velocities.  Plotted for each velocity is 1) each object in the astrometric sample (large histogram) 2) a 30 degree restriction on objects and 3) a 60 degree restriction (smallest histogram).   The 30 and 60 degree restrictions are placed on the X,Y or Z axis and correspond to removing the U,V, or W velocity respectively for objects in cones of noted radius around the respective axis.
 } 
\label{fig:UVW_VEL}
\end{figure}

Figure ~\ref{fig:UVW_VEL} shows our resultant U,V,W distributions where we measure  ($\sigma_{U},\sigma_{V},\sigma_{W})$=(28, 22, 17) km~s$^{-1}$.  We compare these dispersions with the kinematic signatures of the three Galactic populations, namely the thin disk, the thick disk, and the halo.   The overwhelming majority of stars in the solar neighborhood are members of the Galactic disk and these are primarily young thin disk objects as opposed to older thick disk objects.  The halo population of the Galaxy encompasses the oldest population of stars in the Galaxy but these objects are relatively sparse in the vicinity of the Sun.  Membership in any Galactic population has implications for the age and metallicity of the object and kinematics play a large part in defining the various populations.  \citet{2003A&A...398..141S} find ($\sigma_{U},\sigma_{V},\sigma_{W})$=(39 $\pm$ 2, 20 $\pm$ 2, 20 $\pm$ 1) km~s$^{-1}$ for the thin disk and ($\sigma_{U},\sigma_{V},\sigma_{W})$=(63 $\pm$ 6, 39 $\pm$ 4, 39 $\pm$ 4) km~s$^{-1}$ for the thick disk, and \citet{2000AJ....119.2843C} find ($\sigma_{U},\sigma_{V},\sigma_{W})$=(141 $\pm$ 11, 106 $\pm$ 9, 94 $\pm$ 8) km~s$^{-1}$ for the halo portion of the Galaxy.  
Our U,V,W dispersions are consistent (albeit narrower in U ) with the Galactic thin disk.  

\citet{2007ApJ...666.1205Z} -- hereafter Os07-- examined 21 L and T dwarfs and found ($\sigma_{U},\sigma_{V},\sigma_{W})$=(30.2, 16.5, 15.8) km~s$^{-1}$.  Their velocity dispersions are tighter than what is expected from the Galactic thin disk population.  Our calculated dispersions are tighter at U than the Os07 result (which is expected due to the stated bias) but  broader in V and W.    In section 6 we discuss the implications on age of the differences calculated from our astrometric sample.

\subsection{$Red$ $and$ $Blue$ $Photometric$ $Outliers$}
	As discussed in Kirkpatrick et al. (2005) the large number of late-type M, L, and T dwarfs discovered to date has revealed a broad diversity of colors and spectral characteristics, including specific subgroups of peculiar sources that are likely related by their common physical properties.  As a very basic metric, near-IR colors provide one means of distinguishing between  ``normal'' and ``unusual'' objects.  To investigate our sample for kinematically distinct photometric outliers, we first defined the average color ($(J-K_{s})_{avg}$) as well as standard deviation ($\sigma_{J-K_{s}}$) as a function of spectral type using all known ultracool dwarfs (i.e., both with and without proper motion measurements).   Defining the $(J-K_{s})_{avg}$ for spectral bins has been done in previous ultracool dwarf studies such as Kirkpatrick et al. (2000), Vrba et al. (2004), and \citet{2008AJ....135..785W}  but we have included all ultracool dwarfs in the dwarfarchives complilation and the updated photometry reported in Chiu et al. (2006, 2008) and Knapp et al. (2004) which we have converted from the MKO system to the 2MASS system.  Objects were eliminated from the photometric sample if they fit any of the following criteria:
	\begin{enumerate}

	\item  Uncertainty in J or $K_{s}$ $>$  0.2 magnitude;
	\item  Known subdwarf;
	\item  Known binaries unresolved by wide-field imaging surveys  (i.e. separations $\lesssim$ 1 $\arcsec$ e.g. \citealt{1999Sci...283.1718M}; \citealt{2003AJ....126.1526B}; \citealt{2006ApJS..166..585B}; \citealt{2003ApJ...587..407C}; \citealt{2006ApJ...647.1393L}; \citealt{2006AJ....132..891R}); and
	\item  Member of a star forming region (such as Orion) or open cluster (such as the Pleiades)  indicating an age $\lesssim$ 100 Myr (e.g. \citealt{2007ApJ...657..511A}; \citealt{2002ApJ...578..536Z})
\end{enumerate}
We then designated objects as photometric outliers if they satisfied the following criterion :
\begin{equation}
\Delta_{J-K_{s}}=\mid{(J-K_{s})-(J-K_{s})_{avg}}\mid\geq{\rm max}(2\sigma_{J-K_{s}},0.4)
\end{equation}

In other words, if an object's $J-K_{s}$ color was more than twice the standard deviation of the color range for that spectral bin than we flagged it as a red or blue photometric outlier.  If twice the standard deviation was larger than 0.4 magnitudes then it was automatically reset to 0.4. We chose 0.4 as the maximum upper limit for $2\sigma_{J-K_{s}}$ as this is the $\Delta_{J-K_{s}}$ for the entire ultracool dwarf population.  

\begin{figure*}[htbp]
\begin{centering}
\includegraphics[width=1.0\hsize]{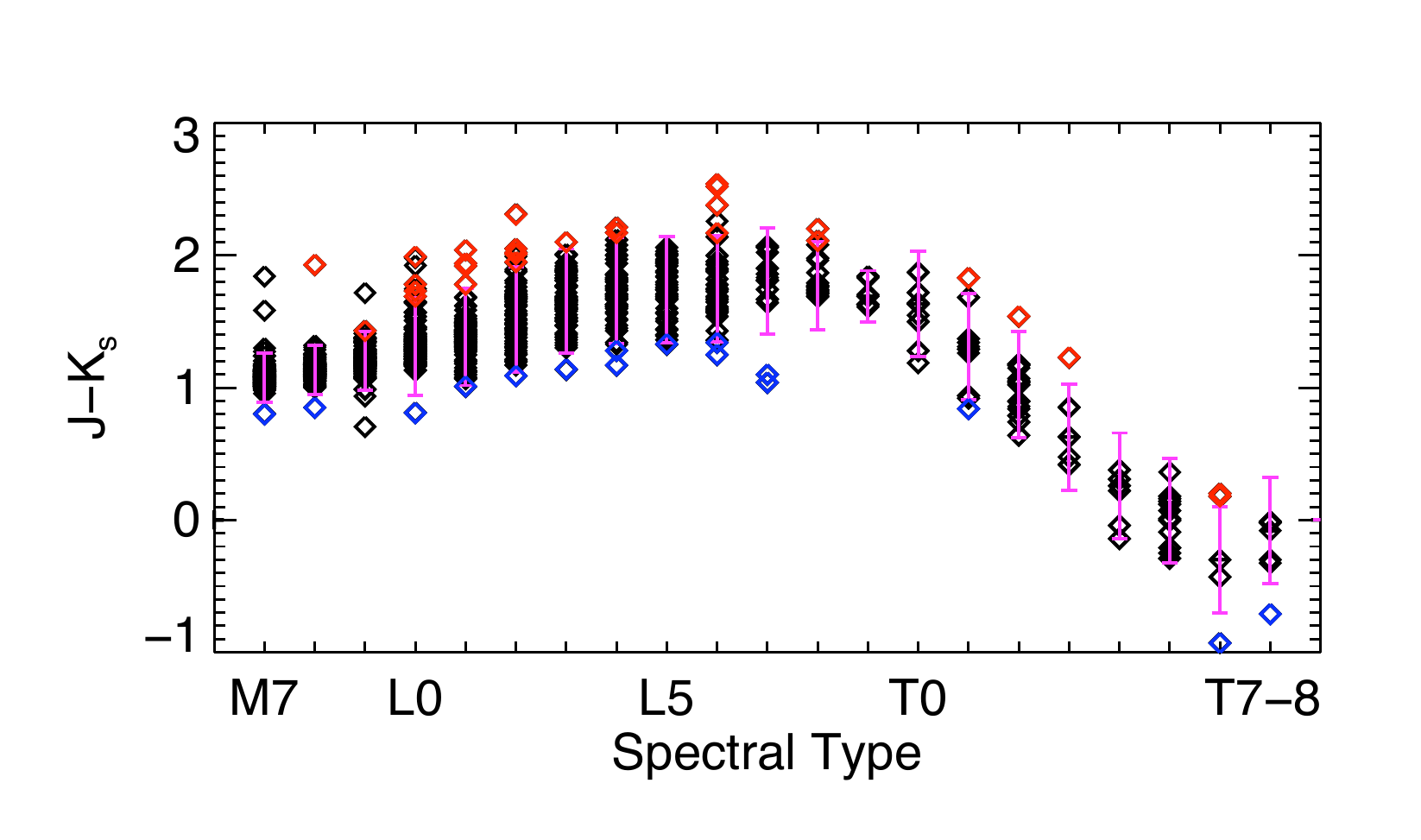}
\end{centering}
\caption{ $J - K_{s}$ colors of late-type dwarfs.  We compute the average values for each spectral type (binned by 1 subtype) from the 2MASS photometry of a select sample of dwarfs and then flag objects as photometric outliers when they are either twice the standard deviation of $J -K_{s}$ or 0.4 magnitude redder or bluer than the average value.  Red symbols above the plotted range of $J - K_{s}$ colors are the red outliers and blue symbols below the plotted range of $J - K_{s}$ colors are the blue outliers.
 } 
\label{fig:JMK_OUT}
\end{figure*}

There are relatively few objects in each spectral bin beyond L9. For spectral type (SpT) $<$ L9  there is a mean of 45 objects used per bin whereas for SpT$>$ L9 there is a mean of only 7 objects.  So photometric outliers are difficult to define for the lower temperature classes and may contaminate the analysis. We grouped T7-T8 dwarfs to improve the statistics used to calculate average values.  The kinematic results for this subset of the ultracool dwarf population are  reported with and without the T dwarfs in Table 8.   Figure ~\ref{fig:JMK_OUT} shows the resulting $J-K_{s}$ color distribution and highlights the photometric outliers. Tables 9 - 10 list the details of the red and blue photometric outliers respectively. Table 5 lists the resultant mean photometric values for each spectral type.  
\\ \indent 

\begin{figure}[!ht]
\begin{center}
\epsscale{1.5}
\plottwo{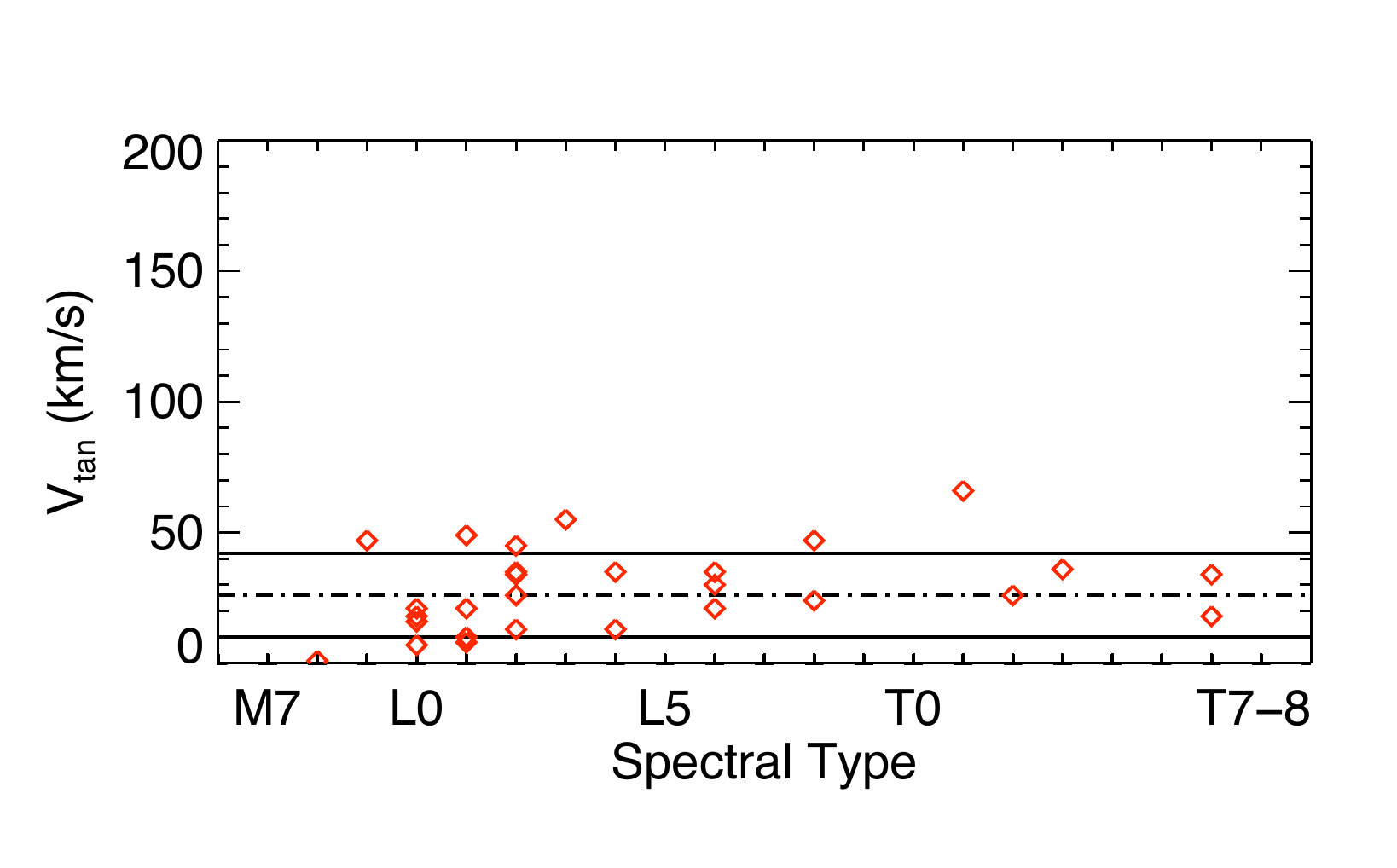}{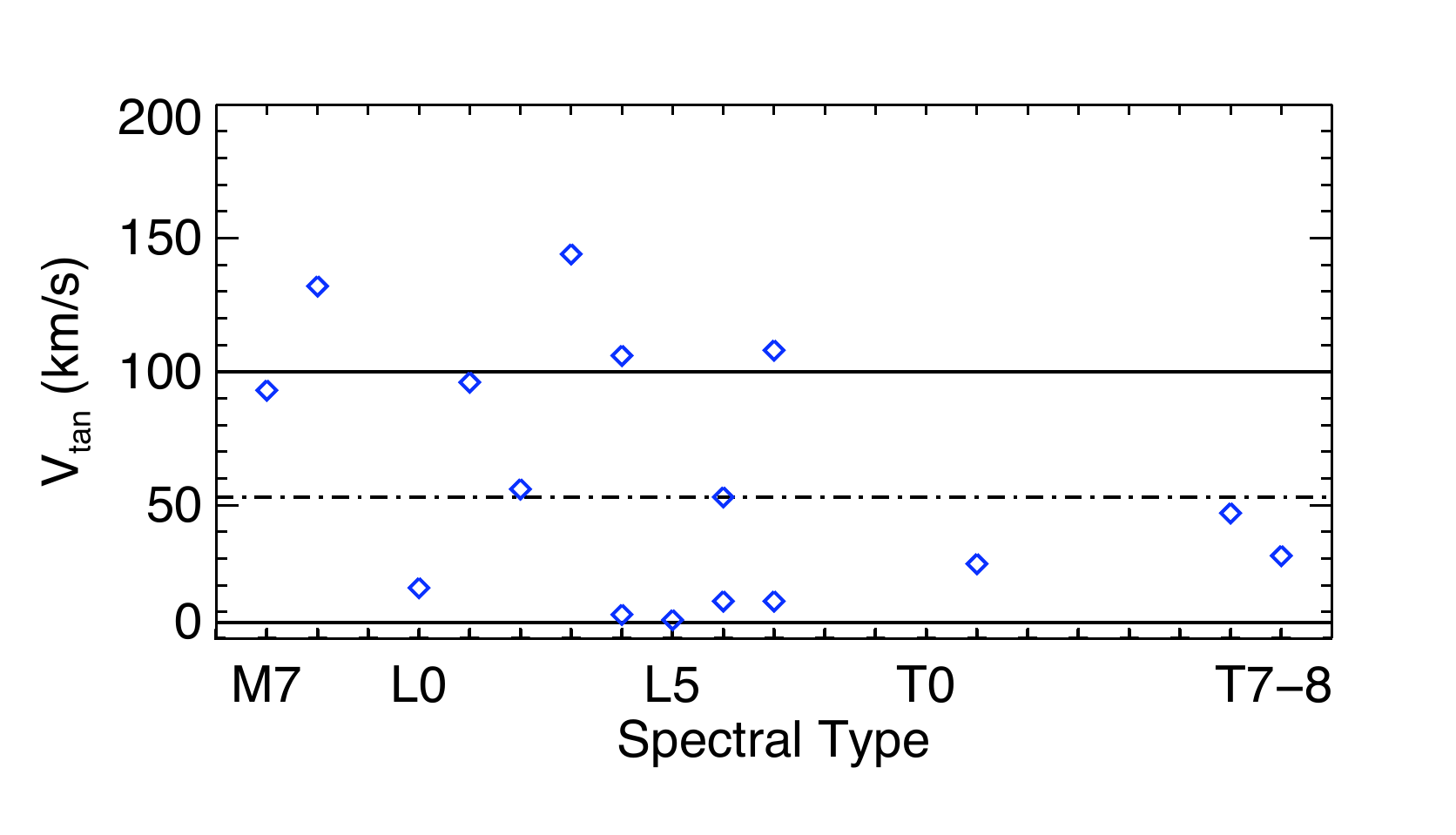}
\end{center}
\caption{ The spread of tangential velocities for objects marked as red outliers (top panel) and blue outliers (bottom panel).  The red population has a fairly tight dispersion and the blue population has a fairly wide dispersion compared to the full sample suggesting a link between near-IR color and age.  The dashed line in each plot represents the median $V_{tan}$ value for the outlier group and the solid black lines represent the dispersion.
 } 
\label{fig:RED_BLUE}
\end{figure}

\begin{figure*}[htbp]
\begin{centering}
\includegraphics[width=1.0\hsize]{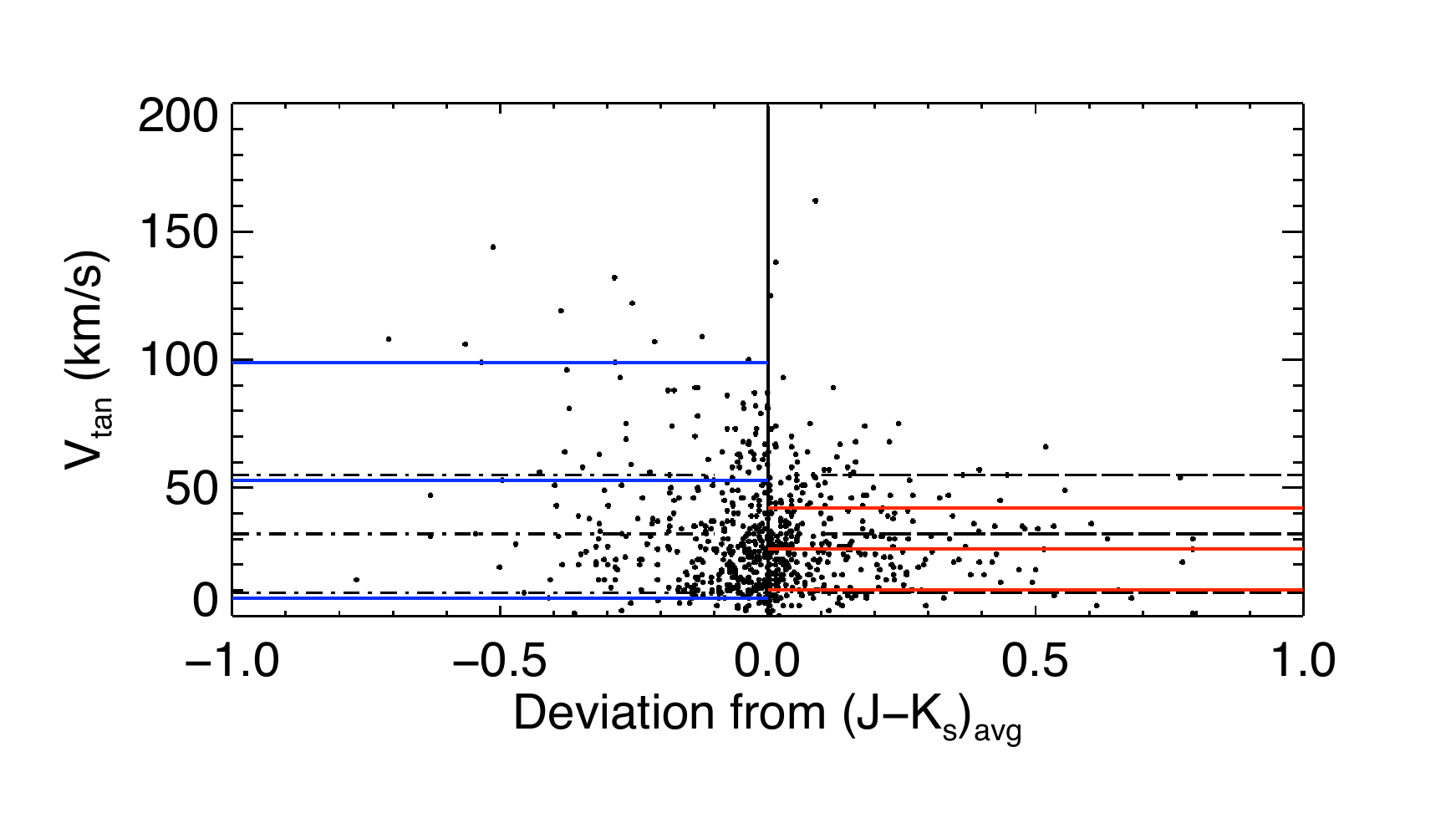}
\end{centering}
\caption{ A scatter plot showing $V_{tan}$ as a function of the deviation in $J-K_{s}$ color from the average at a given spectral type.  The blue outliers appear to move faster on average than the red outliers.  To demonstrate this we have over-plotted the average $V_{tan}$ with dispersion for the blue and red photometric outliers as well as for the full astrometric sample (dashed lines).
 } 
\label{fig:JMK_Scat}
\end{figure*}

Amongst the full sample, we find 16 blue photometric outliers and 29 red photometric outliers.  Many of the objects have already been noted in the literature as having unusual colors, and several of these have anomalous spectra and have been analyzed in detail (e.g. \citealt{2007arXiv0710.1123B}, \citealt{2004AJ....127.3553K}, \citealt{2007MNRAS.378..901F}, \citealt{2006AJ....131.2722C}).  Table 8  lists the mean kinematic properties for the blue and red subgroups of the ultracool dwarf population and Figure ~\ref{fig:RED_BLUE} isolates the outliers and plots their tangential velocity vs. spectral type.  The blue outliers have a median $V_{tan}$ value of 53 km~s$^{-1}$ and a $\sigma_{tan}$ of 47 km~s$^{-1}$ while the red outliers have a median $V_{tan}$ value of 26 km~s$^{-1}$ and a $\sigma_{tan}$ of 16 km~s$^{-1}$ .  Figure ~\ref{fig:JMK_Scat} shows the  tangential velocity vs. J-$K_{s}$ deviation for all objects in the sample with the dispersions of the red and blue outliers highlighted.  There is a clear trend for $V_{tan}$ values to decrease from objects that are blue for their spectral type to those that are red.  This is particularly significant at the extreme edges of this diagram.  The dashed line in Figure ~\ref{fig:JMK_Scat} marks the spread of $V_{tan}$ values for the full sample and demonstrates the significant deviations for the color outliers.   We explore the age differences from these measurements in section 6. 

\subsection{$Low$ $Gravity$ $Objects$}
A number of ultracool dwarfs that exhibit low surface-gravity features have been reported in the literature within the past few years (e.g. \citealt{2007AJ....133..439C}, \citealt{1999ApJ...525..440L}; \citealt{2004ApJ...600.1020M},  \citealt{2006ApJ...639.1120K}; \citealt{2007ApJ...657..511A}).  Low surface-gravity dwarfs are distinguished as such by the presence of weak alkali spectral features, enhanced metal oxide absorption, and reduced $H_{2}$ absorption.  They are most likely young , with lower masses than older objects of the same spectral type.  For ages $\lesssim $ 100 Myr these objects may also have larger radii than older brown dwarfs and low mass stars with similar spectral types, as they are still contracting to their final radii (e.g., \citealt{1997ASPC..119....9B}).   

We examine the kinematics of 37 low surface-gravity dwarfs in this paper. Seven of these objects are flagged as red photometric outliers and were examined in the previous subsection.  The overlap between these two subgroups is not surprising as the reduced $H_{2}$ absorption in low surface-gravity dwarfs leads to a redder near-IR color.  The median $V_{tan}$ value for this subgroup is 18 km~s$^{-1}$ and the $\sigma_{tan}$ value is 15 km~s$^{-1}$ which is smaller than  that of the red photometric outliers as a whole and therefore points to the same conclusion.   The smaller median $V_{tan}$ and tighter dispersion of the low surface-gravity dwarfs as compared to either the full or 20 pc sample indicates that they are kinematically distinct.  

\subsection{$Unusually$ $Blue$ $L$ $Dwarfs$}
A subgroup of unusually blue L dwarfs (UBLs) has been  distinguished based on strong near-IR H$_2$O, FeH and \ion{K}{1} spectral features but otherwise normal optical spectra.  \citet{2008ApJ...674..451B} --hereafter B08-- identify ten objects that comprise this subgroup (see Table 6 in B08).  With the kinematics reported in this article we are able to analyze all ten.  There are several physical mechanisms that can contribute to the spectral properties of UBLs.  High surface-gravity, low metallicity, thin clouds or unaccounted multiplicity are amongst the physical mechanisms most often cited.  B08 has demonstrated that while sub-solar metallicity and high surface-gravity could be contributing factors in explaining the spectral deviations, thin, patchy or large-grained condensate clouds at the photosphere appears to be the primary cause for the anomalous near-IR spectra (e.g. \citealt{2001ApJ...556..872A}, \citealt{2006ApJ...640.1063B}).  
	
The median $V_{tan}$  value for this subgroup is 99 km~s$^{-1}$ with $\sigma_{tan}$ of 47 km~s$^{-1}$ and this subgroup consists of dwarfs with the largest $V_{tan}$ values measured in this kinematic study.  These kinematic results strengthen the case that the UBLs represent an older population and that the blue near-IR colors and spectroscopic properties of these objects are  influenced  by large surface-gravity and/or slightly subsolar metallicities.  Both of these effects may be underlying explanations for the thin clouds seen in blue L dwarf photospheres. Subsolar metallicity reduces the elemental reservoir for condensate grains while high surface-gravity may enhance gravitational settling of clouds.  In effect, the clouds of L dwarfs may be tracers of their age and/or metallicity.   

Eight of the ten UBLs examined in this subsection are also flagged as blue photometric outliers and examined in detail above.  The overlap between these two subgroups is not surprising as many of the UBLs were initially identified by their blue near-IR color (e.g. Cruz et al. 2007, Knapp et al. 2004).  There are 8 other blue photometric outliers, one of which has a $V_{tan}$ value exceeding 100 km~s$^{-1}$.  We plan on obtaining near-IR spectra for these outliers to investigate the possibility that they exhibit similar near-IR spectral features to the UBLs.

While the UBLs are the most kinematically distinct subgroup analyzed in this paper, their kinematics do not match those of the ultracool subdwarfs.  The subdwarfs were excluded from the kinematic analysis in this paper because they are confirmed members of a separate population.  The median $V_{tan}$ value for this subgroup is 196 km~s$^{-1}$ with $\sigma_{tan}$ of 91 km~s$^{-1}$. The UBLs move at half this speed indicating there is a further distinction between UBLs and the metal-poor halo population of ultracool dwarfs.  

\section{$High$ $Velocity$ $Dwarfs$}
Table 11 summarizes the properties of the 14 high velocity dwarfs whose $V_{tan}$ measurements exceed 100 km~s$^{-1}$.  A number of these have been discussed in the literature, having been singled out in their corresponding discovery papers as potential members of the thick disk or halo population. One high velocity dwarf is presented here for the first time.   SDSS J093109.56+032732.5 is an L7.5 dwarf and is classified as both a UBL and a blue photometric outlier.  We calculate $V_{tan}$ for this object to be 108 $\pm$ 23 km~s$^{-1}$.   

Among the high velocity dwarfs, 11 have colors that are  blue and 3 have colors that are normal for their spectral type. Three objects belong to the UBL subgroup.  Three of the objects are late-type M dwarfs (2MASS J18261131+3014201, 2MASS J03341218-4953322, and 2MASS J132352+301433), one is a late T7.5 dwarf (2MASSJ 11145133-2618235) and the rest are early to mid-type L dwarfs.   Four of the objects are flagged as blue photometric outliers. We explore the possibility that these objects are thick disk or halo objects in detail in a forthcoming paper.

 \section{ON THE AGES OF THE ULTRACOOL DWARF POPULATIONS}
 \subsection{$Kinematics$ $and$ $Ages$}
 
 A comparison of the velocity dispersion for nearby stellar populations can be an indicator of age. While individual $V_{tan}$ measurements cannot provide individual age determinations due to scatter and projection effects, the random motions of a population of disk stars are known to increase with age.  This effect is known as the disk age-velocity relation (AVR) and is simulated by fitting well-constrained data against the following analytic form:
\begin{equation}
        \sigma(t)=\sigma_0(1+\frac{t}{\tau})^\alpha
\end{equation}
where $\sigma(t)$ is the total velocity dispersion as a function of time, $\sigma_0$ is the initial velocity dispersion at t=0, $\tau$ is a constant with unit of time, and $\alpha$ is the heating index (\citealt{1977A&A....60..263W}).   $\sigma(t)$ is defined for U,V,W space velocities but we can estimate the total velocity dispersion using our measured tangential velocities assuming the dispersions are spread equally between all three velocity components, such that:
\begin{equation}
        \sigma(t)=(3/2)^{(1/2)}\sigma_{tan}
\end{equation}

\citet{2002MNRAS.337..731H} calculate $\alpha$ from seven distinct data sets (both pre and post- Hipparcos) and find $\alpha$ ranges from 0.3 to 0.6.  This is a large range of values and the authors are reluctant to assign a higher likelihood to any given value as each have nearly equal uncertainties.   One possible explanation for the spread of values is that $\sigma$ should be  mass dependent \footnote{Indeed \citealt{1980Ap&SS..73..435I} proposes the introduction of a mass term to account for the importance of the exchange of energy between stars in the Galactic disk}.  If so, this would make a large difference in the age calculations for the low mass ultracool dwarf population.  While the age-velocity relation in the nearby disk remains only roughly determined, there is strong observational evidence for a relation so we proceed with caution in examining the broad age possibilities implied by the AVR for the ultracool dwarf population. 
 
\begin{figure}[!ht]
\begin{center}
\epsscale{1.5}
\plottwo{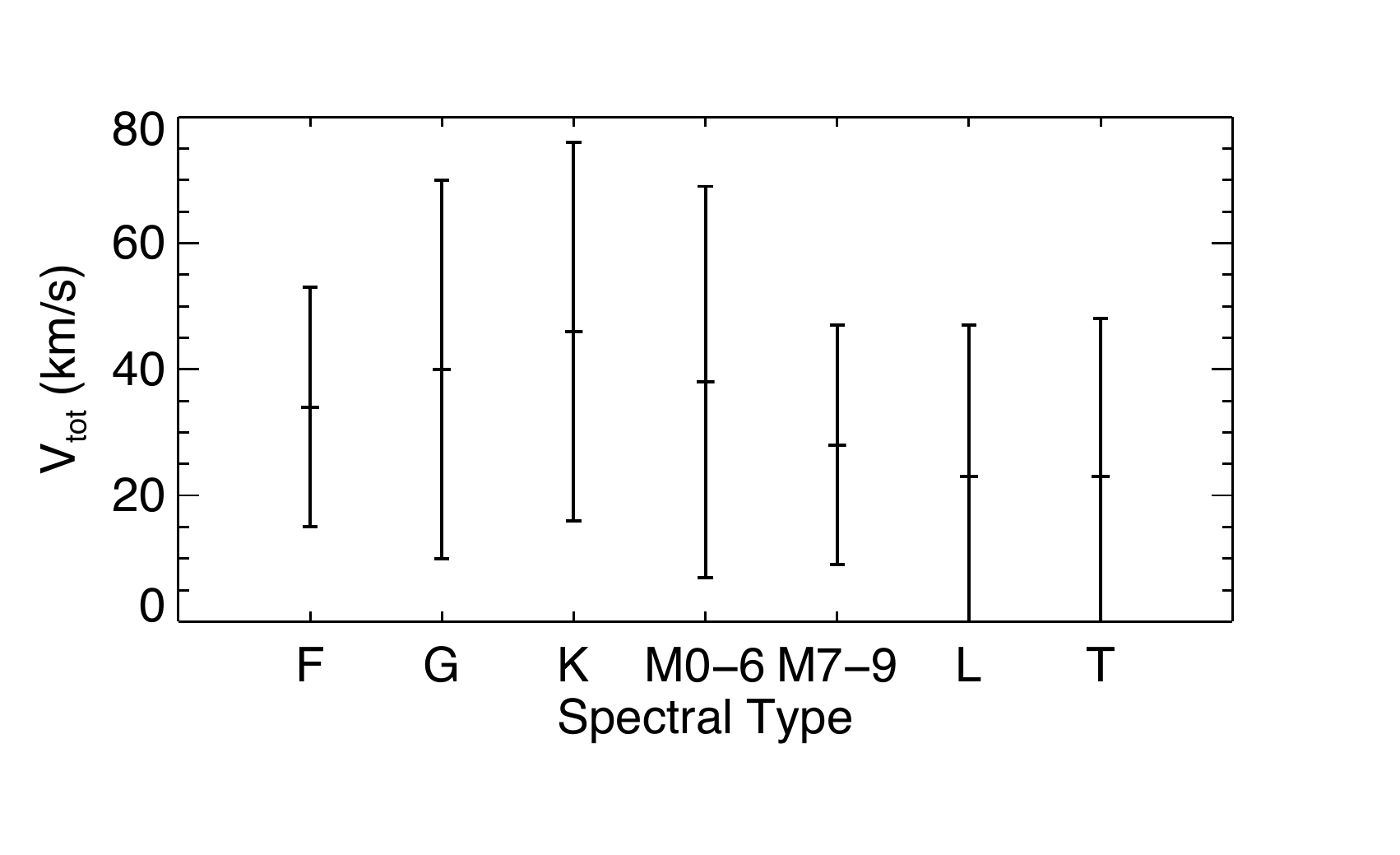}{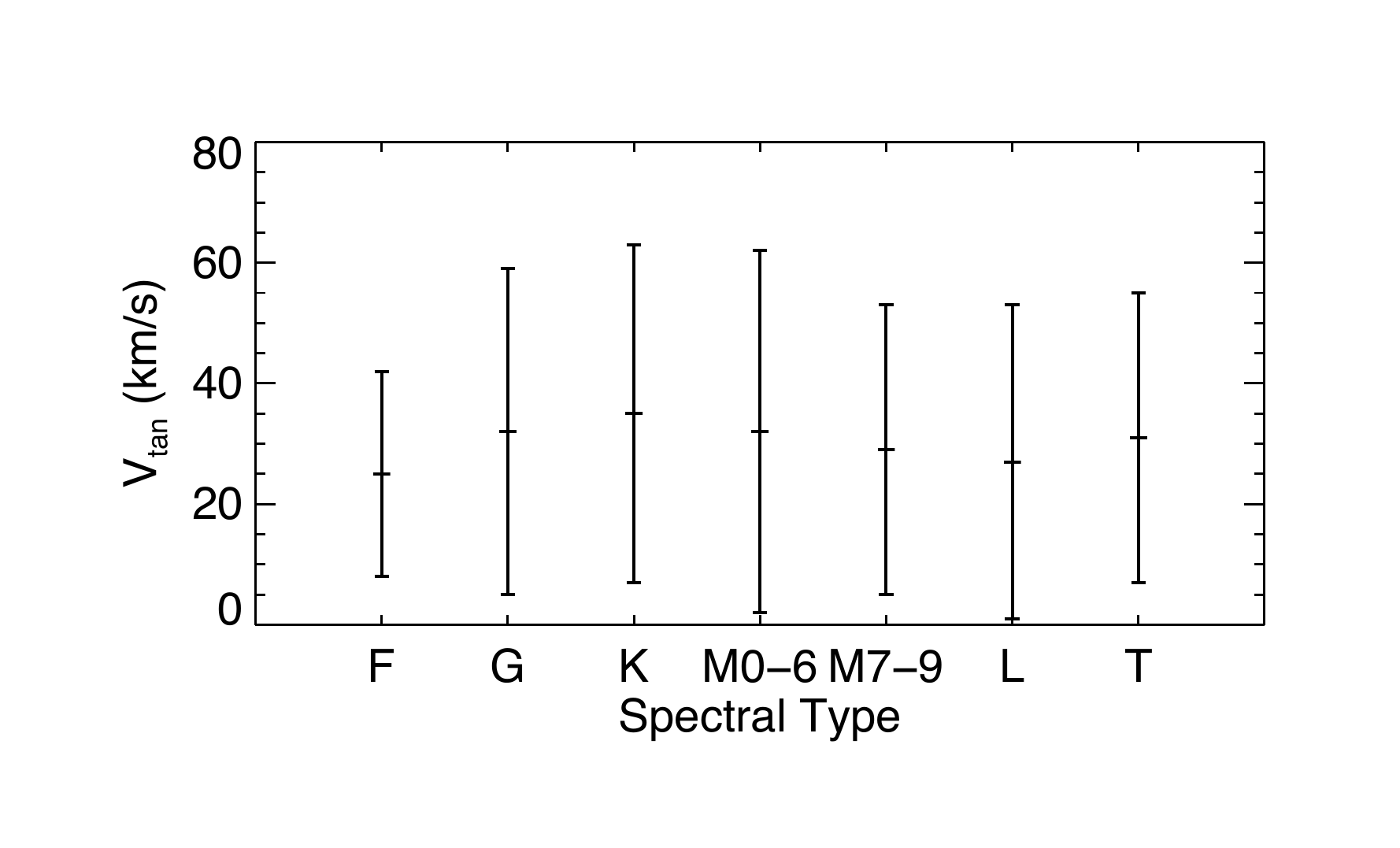}
\end{center}
\caption{ Top: A plot of median $V_{tot}$ and $\sigma_{tot}$ values calculated from the U,V,W velocities for the 20 pc sample of F through T  objects.  Bottom: A plot of median $V_{tan}$ and $\sigma_{tan}$ values calculated from the proper motions and distances  for the 20 pc sample of F through T  objects.} 
\label{fig:VTAN_Spread}
\end{figure}

Recent findings have suggested that late-type M stars in the solar neighborhood are younger on average than earlier type stars (\citealt{1988MNRAS.234..177H}; \citealt{1994ApJS...94..749K}; \citealt{1994AJ....108.1456R}).  Several investigators have combined kinematics with the \citet{1977A&A....60..263W} relationship (which uses a value of 1/3 for $\alpha$) to estimate age ranges for the ultracool dwarf population and concluded that it is kinematically younger than nearby stellar populations (e.g. \citealt{2002AJ....124.1170D}, \citealt{2007AJ....133.2258S}, \citealt{2000AJ....120.1085G}, \citealt{2007ApJ...666.1205Z}).  We conducted a direct $V_{tan}$ comparison with nearby stellar populations to draw conclusions about the kinematic distinguishability of our ultracool dwarf sample.  We compared the kinematics of a 20 pc sample of F,G,K, and early M stars from \citet{2003A&A...398..141S}, \citet{2004yCat.3239....0K} and \citet{2004A&A...418..989N} using a limiting proper motion of 25 mas/yr to our 20 pc sample and examined the resultant median $V_{tan}$, $\sigma_{tan}$, and $V_{tot}$ , $\sigma_{tot}$  values (where $V_{tot}$ comes from the U,V,W velocities).  Figure ~\ref{fig:VTAN_Spread} shows our resultant velocity dispersions for nearby stellar populations along with the dispersions of our 20 pc sample.  We show both the dispersions calculated using tangential velocities and those calculated using U,V,W velocities.  As expected the dispersions are tighter for the UCDs when U,V,W values are used since we have attempted to minimize the importance of radial velocity.  This effect is also reflected in the younger ages estimated from these dispersions.  The tangential velocity dispersions are in good agreement between the UCDs and nearby stellar populations (see also \citealt{2008AJ....135..785W},\citealt{2007AJ....134.2418B}, and  \citet{2008arXiv0807.2452C}). Table 12 contains the calculated kinematic measurements and Wielen ages using both $\sigma_{tan}$ and $\sigma_{tot}$. With Wielen ages of 3-8 Gyr calculated from $\sigma_{tan}$, we conclude that our 20 pc sample is kinematically indistinct from other nearby stellar populations and hence is not kinematically younger.  The ages calculated by the AVR for the 20 pc sample are in good agreement with those predicted in population synthesis models where the mean ages for the ultracool dwarf population range from 3-6 Gyr (\citealt{2004ApJS..155..191B}; \citealt{2005ApJ...625..385A}).  

We do find younger ages for the ultracool dwarf population when the high velocity dwarfs are excluded.   As stated in section 4, the median $V_{tan}$ and $\sigma_{tan}$ values are naturally reduced when the high velocity dwarfs are excluded and consequently the ages are also reduced.  Kinematic analyses of the past have regarded these objects as a separate older population and omitted them from an age calculation (e.g. \citealt{2007AJ....133.2258S}).  Table 6 presents the ages with and without the high velocity dwarfs for the 20 pc sample.  When the high velocity dwarfs are excluded the age ranges are reduced from 3-8 Gyr to 2-4 Gyr, which is still consistent with population synthesis models.  

The Os07 study estimated mean ages of $\sim$ 1 Gyr for the L and T dwarf population .  Even with the exclusion of the kinematic outliers, the ages calculated in our full and 20 pc samples do not match this very young age.  Os07 combined proper motions, precise parallaxes, and radial velocities to study the 3D kinematics of a limited sample of 21 objects.  When we apply an age velocity relation to the red photometric outliers and the low gravity dwarfs we do find ages that are on the order of $\sim$ 1 Gyr.  We discuss the red outliers below but conclude that the low surface-gravity dwarfs are kinematically younger than the full or 20 pc sample.  This result is consistent with what has already been reported through spectroscopic studies.   There do not appear to be any low surface-gravity dwarfs flagged in the Os07 sample however further examination of their L and T dwarf spectra might be warranted by the discrepancy in ages between our samples.  We suggest that kinematic studies of UCDs to date, including Os07, may have been plagued by small number statistics or a bias in the sample analyzed.  

\subsection{$Ages$ $of$ $the$ $Red$ $and$ $Blue$ $Outliers$}
We have defined two subgroups of the ultracool dwarf population in this article that are both photometrically and kinematically distinct from the full or 20 pc samples.  Objects whose $J-K_{s}$ colors are sufficiently deviant are also kinematically different from the overall population.  While we again advise caution in using the AVR, we can use it to compare the predicted ages of the photometric outliers to the predicted ages for the full or 20 pc samples. We find that the kinematics for the red outliers are consistent with a younger population of ultracool dwarfs whereas the kinematics for the blue outliers are consistent with an older population.  The $\sim$ 1 Gyr mean age for the red outliers coincides with the prediction of Os07 for the entire L and T population.  We have examined the photometry of their sample and concluded that the $J-K_{s}$ colors of their objects are normal so the age calculation of their sample is not influenced by a bias from inclusion of photometric outliers.  The $\sim$ 38 Gyr mean age for the blue outliers is misleading. It indicates a large divergence from the full and 20 pc samples but also indicates that the AVR must be incorrect for these objects.  The more informative number in this case is the  median $V_{tan}$ which, at 56 km/s, is nearly twice the expected value for the thin disk (see \citealt{2005nlds.book.....R}).  The blue photometric outliers most likely belong to an older population of the Galaxy such as the thick disk or the halo.   The Wielen AVR is only valid for thin disk objects and we are unaware of an equivalent age relation for the halo or thick disk population.

From our kinematic analysis we conclude that there is an age-color relation that can be derived for the UCD field population.   A change in broad-band collision-induced $H_{2}$ absorption that suppresses flux at K-band  is partially responsible for the near-IR color and consequently the age of the photometric outliers (\citealt{1969ApJ...156..989L}; \citealt{1994ApJ...424..333S}; \citealt{1997A&A...324..185B}).  $H_{2}$ absorption is pressure and hence gravity sensitive. Changes in  $H_{2}$ absorption effect gravity sensitive features which are used as an indicator of age. The overlap of red photometric outliers with low surface-gravity dwarfs and the concensus within the literature that low-g dwarfs are young demonstrates the age sensitivity of $H_{2}$ absorption.  

Cloud properties have also been linked to a change in near-IR color. The analyses of B08, and \citealt{2008ApJ...678.1372C} have shown that the thickness of patchy or large-grained condensate clouds at the photospheres of dwarfs will lead to redder (thick clouds) or bluer (thin clouds) near-IR colors.  The old age implied by the kinematics of the blue outliers and the overlap with the UBLs suggests that there is a correlation between cloud properties and age or metallicity; but further investigation is warranted in this area. 

\citet{2008MNRAS.385.1771J} have proposed a relation for inferring the ages of young L dwarfs using only near-infrared photometry and estimated distances.  Their work supports the argument for an age-color relation for the ultracool dwarf population.  The ages that they work with are no larger than $\sim$ 0.7 Gyr.   At these young ages, the surface gravities of UCDs change more rapidly than for ages greater than a few Gyr, so the age-color relation may be much stronger in the \citet{2008MNRAS.385.1771J} sample than that seen for field dwarfs.  

\section{CONCLUSIONS}

We present new proper motions for 427 late-type M, L, and T dwarfs and combine all previous proper motion measurements with either parallax measurements or spectrophotometric distances  to compute tangential velocities for 841 M7--T9 dwarfs.  We derive average kinematic and photometric values for individual spectral types as well as for the late-type M,L, and  T populations as a whole.  We conduct a crude U,V,W analysis and find that the full and 20 pc samples examined in this article have space velocities consistent with the Galactic thin disk population.  However, there are 14 objects in the ultracool dwarf population that lie at the tail end of the velocity distribution and are likely to be part of an older Galactic population.  Ages for the 20 pc sample of this kinematic study are consistent with the 3-6 Gyr values derived in population synthesis models;  we propose that one reason for prior kinematic reports of $\sim$1 Gyr mean ages for the L and T dwarf population is due to small number statistics or a bias in the sample analyzed. 

We find a large difference in the kinematics between red and blue photometric outliers and conclude that their velocity dispersions are kinematically distinct from the full or 20 pc samples.  Analysis of the low surface-gravity and UBL subgroups also shows a distinction from the full and 20 pc samples.  Applying an age-velocity relation we conclude that the red outliers and low surface-gravity subgroups are younger than the full and 20 pc samples and the blue outliers and UBLs are older.   

\acknowledgments{ We acknowledge receipt of observation time through the SMARTS consortium and MDM consortium.  We especially thank the observing staff at CTIO, including J. Velazquez the night assistant for a week of CPAPIR observing time in January 2008.  Stony Brook's participation in the SMARTS consortium is made possible by generous support by the Dean of Arts and Sciences, the Provost, and the Vice-President for Research of Stony Brook University.  J. Faherty would like to thank S. Lepine for his useful conversations about kinematic studies.   We also thank R. Doyon, E. Artigau, and L. Malo for help with CPAPIR usage and data reduction and K. Schlesinger, R. Assef, and D. Atlee for help with using TIFKAM and the 1.3m telescope at MDM.  Research has benefitted from the M, L, and T dwarf compendium housed at DwarfArchives.org and maintained by Chris Gelino, Davy Kirkpatrick, and Adam Burgasser.  This publication has made use of the VLM Binaries Archive maintained by Nick Siegler at  http://www.vlmbinaries.org. This publication makes use of data products from the Two Micron All Sky Survey, which is a joint project of the University of Massachusetts and the Infrared Processing and Analysis Center/California Institute of Technology, funded by the National Aeronautics and Space Administration and the National Science Foundation.  This research has made use of the NASA/ IPAC Infrared Science Archive, which is operated by the Jet Propulsion Laboratory, California Institute of Technology, under contract with the National Aeronautics and Space Administration.
 }

\bibliographystyle{apj}
\bibliography{paper}

\pagebreak

\pagebreak

\pagebreak

\pagebreak

\pagebreak

\pagebreak

\pagebreak

\pagebreak

\pagebreak

\pagebreak

\pagebreak

\begin{deluxetable}{rrcccccr}
\label{tab:tab1}
\tabletypesize{\scriptsize}
\tablecaption{ Properties of Instruments Used for Astrometric Measurements}
\tablewidth{0pt}
\tablehead{
\colhead{Telescope} &
\colhead{Instrument} &
\colhead{Band} &
\colhead{FOV} &
\colhead{Plate Scale} &
\colhead{Dates} &
\colhead{ Seeing} &
\colhead{Sources Observed} \\
& & &
\colhead{(arcminutes)} &
\colhead{(arcsecond/pixel)} & &
\colhead{(arcseconds)} \\

}
\startdata
        CTIO 0.9m &   CFIM  &      I & 4.5   &       0.40     &  2007 Sep 23 - 26   & 0.8 - 2.0 &  42\\ 
    MDM 1.3m&   TIFKAM  &  J &  5.6    &       0.55  &       2007 Nov 20 - 24   & 1.0 - 2.5 &  66\\ 
       &   &       &  &  & 2008 Apr 22 - 28   & 0.8 - 2.5  &  80\\ 
      CTIO 1.5m&   CPAPIR  & J &     35.0   &       1.02  &       2005 Oct 19 &1.0 - 1.5 &      4          \\ 
      &     &  & &    &       2006 Aug 21  & 1.0 - 2.0  &7 \\ 
      &     &  & &    &       2007 Mar 04  &  0.9 - 1.8  &28\\ 
      &     &  & &    &       2007 Mar 23  &  0.9 - 1.3  &39\\ 
      &     &  & &    &       2007 Dec 03 - 06  & 1.0 - 2.0 &  35\\ 
      &     &  & &    &       2008 Jan 15 - 23  &   0.7 - 2.5 & 248\\ 

      \hline
      \hline
     \enddata
\end{deluxetable}
\pagebreak

\begin{deluxetable}{llllccccrrrrr}
\rotate
\label{tab:tab7}
\tabletypesize{\tiny}
\tablecaption{ New Proper Motion Measurements}
\tablewidth{0pt}
\small
\tablehead{
\colhead{Source Name} &
\colhead{RA} &
\colhead{DEC} &
\colhead{SpT\tablenotemark{a}} &
\colhead{ SpT} &
\colhead{$\mu_{\alpha}cos(\delta)$ } &
\colhead{$ \mu_{\delta}$}  &
\colhead{Baseline} &
\colhead{Instrument}  \\

 & (J2000) & (J2000) &
\colhead{(optical)} &
\colhead{(near-IR)} &
\colhead{($\arcsec$/yr)} &
\colhead{($\arcsec$/yr)} &
\colhead{(yrs)} &  \\
\colhead{(1)} &
\colhead{(2)} &
\colhead{(3)} &
\colhead{(4)} &
\colhead{(5)} &
\colhead{(6)} &
\colhead{(7)} &
\colhead{(8)} &
\colhead{(9)} \\

}
\startdata
2MASS J00034227-2822410	&	00 03 42.27	&	-28 22 41.0	&	M7.5	&	---	&	0.257	$\pm$	0.016	&	-0.145	$\pm$	0.018	&	9.2	&	CPAPIR	\\
2MASSI J0006205-172051	&	00 06 20.50	&	-17 20 50.6	&	L2.5	&	---	&	-0.032	$\pm$	0.017	&	0.017	$\pm$	0.018	&	9.5	&	CPAPIR	\\
2MASS J00100009-2031122	&	00 10 00.09	&	-20 31 12.2	&	L0	&	---	&	0.100	$\pm$	0.022	&	0.007	$\pm$	0.023	&	9.1	&	CFIM	\\
2MASSI J0013578-223520	&	00 13 57.79	&	-22 35 20.0	&	L4	&	---	&	0.055	$\pm$	0.017	&	-0.051	$\pm$	0.019	&	9.4	&	CPAPIR	\\
2MASS J00145575-4844171	&	00 14 55.75	&	-48 44 17.1	&	L2.5	&	---	&	0.851	$\pm$	0.012	&	0.289	$\pm$	0.018	&	8.1	&	CPAPIR	\\
2MASS J00165953-4056541	&	00 16 59.53	&	-40 56 54.1	&	L3.5	&	---	&	0.201	$\pm$	0.014	&	0.032	$\pm$	0.018	&	8.3	&	CPAPIR	\\
EROS-MP J0032-4405	&	00 32 55.84	&	-44 05 05.8	&	L0	&	---	&	0.126	$\pm$	0.015	&	-0.099	$\pm$	0.021	&	8.4	&	CPAPIR	\\
2MASS J00332386-1521309	&	00 33 23.86	&	-15 21 30.9	&	L4	&	---	&	0.291	$\pm$	0.016	&	0.043	$\pm$	0.017	&	8.3	&	CPAPIR	\\
\enddata
\tablecomments{Details on the new proper motion measurements reported in this article. See Table 3 for discovery references.}
\tablenotetext{a}{SpT refers to the spectral type of the object.}
\end{deluxetable}

\begin{deluxetable}{l l l l l l l l l l l l l l l l}
\rotate
\label{tab:tab7}
\tabletypesize{\tiny}
\tablecaption{ Full Astrometric Database}
\tablewidth{0pt}
\small
\tablehead{
\colhead{Source Name} &
\colhead{Ref.} &
\colhead{RA } &
\colhead{DEC} &
\colhead{2MASS J} &
\colhead{2MASS $K_{s}$} &
\colhead{$\mu_{\alpha}cos(\delta)$ } &
\colhead{$ \mu_{\delta}$}  &
\colhead{$\mu$ Ref.} &
\colhead{SpT} &
\colhead{SpT}&
\colhead{Distance}&
\colhead{$V_{tan}$}&
\colhead{Note\tablenotemark{r}}  &
\colhead{Epoch}  \\

 & &
 \colhead{(J2000)} &
 \colhead{(J2000)} &
\colhead{(mag)} &
\colhead{ (mag)} &
\colhead{($\arcsec$/yr)} &
\colhead{(\arcsec/yr)} &
&
\colhead{(opt)} &
\colhead{(IR)} &
\colhead{(pc)} &
\colhead{(km~s$^{-1}$)} \\

\colhead{(1)} &
\colhead{(2)} &
\colhead{(3)} &
\colhead{(4)} &
\colhead{(5)} &
\colhead{(6)} &
\colhead{(7)} &
\colhead{(8)} &
\colhead{(9)} &
\colhead{(10)} &
\colhead{(11)} &
\colhead{(13)}&
\colhead{(14)}&
\colhead{(15)} &
\colhead{(16)} \\


}
\startdata SDSS J000013.54+255418.6	&	54	&	00 00 13.54	&	+25 54 18.0	&	14.99	$\pm$	0.10\tablenotemark{c}	&	14.73	$\pm$	0.06\tablenotemark{c}	&	0.006	$\pm$	0.019	&	0.130	$\pm$	0.022	&	10	&	---	&	T4.5	&	11	$\pm$	1		&	7	$\pm$	1	&	---	&	1998.8	\\
SDSS J000112.18+153535.5	&	54	&	00 01 12.17	&	+15 35 35.5	&	15.42	$\pm$	0.06\tablenotemark{c}	&	13.56	$\pm$	0.10\tablenotemark{c}	&	0.150	$\pm$	0.023	&	-0.169	$\pm$	0.015	&	10	&	---	&	L4	&	29	$\pm$	6		&	31	$\pm$	7	&	---	&	2000.7	\\
2MASS J00034227-2822410	&	20	&	00 03 42.27	&	-28 22 41.0	&	13.07	$\pm$	0.02		&	11.97	$\pm$	0.03		&	0.257	$\pm$	0.016	&	-0.145	$\pm$	0.018	&	19	&	M7.5	&	---	&	26	$\pm$	3		&	37	$\pm$	4	&	VLMC-Possible from Kelle	&	1998.9	\\
GJ 1001B, LHS 102B	&	37	&	00 04 34.84	&	-40 44 05.8	&	13.11	$\pm$	0.02		&	11.40	$\pm$	0.03		&	0.644	$\pm$	0.003	&	-1.494	$\pm$	0.002	&	23	&	L5	&	L4.5	&	13.0	$\pm$	0.7\tablenotemark{j}	&	100.4	$\pm$	5.2	&	CB	&	1999.6	\\
2MASS J00044144-2058298	&	45	&	00 04 41.44	&	-20 58 29.8	&	12.40	$\pm$	0.02		&	11.40	$\pm$	0.02		&	0.826	$\pm$	0.076	&	-0.009	$\pm$	0.075	&	27	&	M8	&	---	&	18	$\pm$	3		&	70	$\pm$	14	&	---	&	1999.5	\\
2MASS J00054844-2157196	&	86	&	00 05 48.44	&	-21 57 19.6	&	13.27	$\pm$	0.03		&	12.20	$\pm$	0.03		&	0.703	$\pm$	0.024	&	-0.119	$\pm$	0.004	&	28	&	M9	&	---	&	23	$\pm$	3		&	78	$\pm$	11	&	---	&	1999.5	\\
2MASSI J0006205-172051	&	43	&	00 06 20.50	&	-17 20 50.6	&	15.66	$\pm$	0.07		&	14.01	$\pm$	0.05		&	-0.032	$\pm$	0.017	&	0.017	$\pm$	0.018	&	19	&	L2.5	&	---	&	43	$\pm$	4		&	7	$\pm$	4	&	---	&	1998.5	\\
2MASS J00070787-2458042	&	86	&	00 07 07.87	&	-24 58 04.2	&	13.12	$\pm$	0.02		&	12.06	$\pm$	0.02		&	0.189	$\pm$	0.022	&	-0.051	$\pm$	0.006	&	28	&	M7	&	---	&	30	$\pm$	4		&	28	$\pm$	5	&	---	&	1998.9	\\
\enddata
\end{deluxetable}

\begin{deluxetable}{r r r r c c c r r r r r|}
\label{tab:tab7}
\tabletypesize{\scriptsize}
\tablecaption{ Discrepant Proper Motion Values}
\tablewidth{0pt}
\tablehead{
\colhead{Name} &
\colhead{$\mu_{\alpha}cos(\delta)$} &
\colhead{$\mu_{dec}$ } &
\colhead{$\mu_{\alpha}cos(\delta)$} &
\colhead{$\mu_{dec}$ } &
\colhead{Reference} \\

 & 
\colhead{($\arcsec$/yr)} &
\colhead{($\arcsec$/yr)} &
\colhead{($\arcsec$/yr)} &
\colhead{($\arcsec$/yr)} \\
&
\colhead{This Paper} &
\colhead{This Paper} &
\colhead{Literature} &
\colhead{Literature}   \\

}
\startdata
SIPS J0050-1538	&	-0.229	$\pm$	0.018	&	-0.494	$\pm$	0.019	&	-0.495	$\pm$	0.039	&	-0.457	$\pm$	0.038	&	16	\\
\hline
2MASSJ02271036-1624479	&	0.426	$\pm$	0.016	&	-0.297	$\pm$	0.017	&	0.509	$\pm$	0.016	&	-0.303	$\pm$	0.010	&	16	\\
\hline
2MASSJ09393548-2448279	&	0.592	$\pm$	0.019	&	-1.064	$\pm$	0.021	&	0.486	$\pm$	0.031	&	-1.042	$\pm$	0.055	&	41	\\
\hline
2MASSWJ1155395-372735	&	0.050	$\pm$	0.012	&	-0.767	$\pm$	0.015	&	0.113	$\pm$	0.005	&	-0.861	$\pm$	0.039	&	16	\\
	&				&				&	0.013	$\pm$	0.015	&	-0.778	$\pm$	0.013	&	9	\\
	&				&				&	0.06	$\pm$	0.04	&	-0.82	$\pm$	0.07	&	36	\\
\hline
2MASSJ13411160-3052505	&	0.030	$\pm$	0.013	&	-0.134	$\pm$	0.015	&	0.109	$\pm$	0.014	&	-0.163	$\pm$	0.022	&	17	\\
\hline
2MASSJ13475911-7610054	&	0.203	$\pm$	0.005	&	0.038	$\pm$	0.020	&	0.257	$\pm$	0.063	&	0.287	$\pm$	0.063	&	22	\\
\enddata
\tablecomments{Details on the discrepant proper motion objects.  We note only objects whose proper motion values were discrepant by more than 2$\sigma$.
Proper motion references are listed in Table 3.}
\end{deluxetable}

\begin{deluxetable}{llcccccccccc}
\label{tab:tab5}
\tabletypesize{\scriptsize}
\tablecaption{ Median Photometric and Kinematic Properties of Ultracool Dwarfs}
\tablewidth{0pt}
\tablehead{
\colhead{SpT} &
\colhead{$N_{\mu}$} &
\colhead{$\mu_{median}$} &
\colhead{$\sigma_{\mu}$} &
\colhead{Median Dist} &
\colhead{ $\sigma_{dist}$} &
\colhead{$N_{J-K_{s}}$} &
\colhead{$(J-K_{s})_{avg}$ } &
\colhead{2*$\sigma_{J-K_{s}}$} &
\colhead{$N_{Red}$} &
\colhead{$N_{Blue}$} \\
 & &
\colhead{($\arcsec$/yr)} &
\colhead{($\arcsec$/yr)} &
\colhead{(pc)} &
\colhead{(pc)}  \\
\colhead{(1)} &
\colhead{(2)} &
\colhead{(3)} &
\colhead{(4)} &
\colhead{(5)} &
\colhead{(6)} &
\colhead{(7)} &
\colhead{(8)} &
\colhead{(9)} &
\colhead{(10)} &
\colhead{(11)} \\
}
\startdata
M7	&	88	&	0.261	&	0.553	&	25	&	10	&	160	&	1.08	&	0.19	&	0	&	1	\\
M8	&	114	&	0.210	&	0.403	&	23	&	8	&	147	&	1.14	&	0.18	&	1	&	1	\\
M9	&	71	&	0.204	&	0.357	&	22	&	10	&	107	&	1.20	&	0.22	&	1	&	0	\\
L0	&	93	&	0.111	&	0.211	&	32	&	19	&	92	&	1.31	&	0.37	&	4	&	1	\\
L1	&	83	&	0.208	&	0.301	&	31	&	21	&	82	&	1.39	&	0.37	&	4	&	1	\\
L2	&	58	&	0.185	&	0.209	&	32	&	17	&	63	&	1.52	&	0.40	&	5	&	1	\\
L3	&	64	&	0.189	&	0.398	&	33	&	17	&	67	&	1.65	&	0.39	&	1	&	1	\\
L4	&	50	&	0.183	&	0.284	&	27	&	12	&	44	&	1.73	&	0.40	&	2	&	2	\\
\enddata
\tablecomments{To calculate the $(J-K_{s})_{avg}$ for each spectral type, we chose only objects that were not identified as binaries,young cluster members, subdwarfs and/or had $\sigma_{J}$ and $\sigma_{K_{s}}$ $<$ 0.20.
}
\end{deluxetable}

\begin{deluxetable}{llccccccc}
\tabletypesize{\scriptsize}
\label{tab:tab1}
\tablecaption{20 pc Sample}
\tablewidth{0pt}
\tablehead{
\colhead{SpT} &
\colhead{N} &
\colhead{N } &
\colhead{median $V_{tan}$} &
\colhead{median $V_{tan}$} &
\colhead{$\sigma_{tan}$} &
\colhead{$\sigma_{tan}$} &
\colhead{Age} &
\colhead{Age} \\
& &
\colhead{High $V_{tan}$}&
&
\colhead{with high $V_{tan}$}&
&
\colhead{with high $V_{tan}$}&
&
\colhead{with high $V_{tan}$}\\
 & & & \colhead{(km~s$^{-1}$)} &
\colhead{(km~s$^{-1}$)} &
\colhead{(km~s$^{-1}$)} &
\colhead{(km~s$^{-1}$)} &
\colhead{(Gyr)} &
\colhead{(Gyr)} \\
\colhead{(1)} &
\colhead{(2)} &
\colhead{(3)} &
\colhead{(4)} &
\colhead{(5)} &
\colhead{(6)} &
\colhead{(7)} &
\colhead{(8)} &
\colhead{(9)} \\
}
\startdata
M7	&	29	&	0	&	25	&	25	&	20	&	20	&	---	&	---	\\
M8	&	37	&	1	&	33	&	33	&	20	&	25	&	---	&	---	\\
M9	&	27	&	1	&	26	&	26	&	22	&	26	&	---	&	---	\\
L0	&	9	&	0	&	19	&	19	&	21	&	21	&	---	&	---	\\
L1	&	19	&	0	&	30	&	30	&	29	&	29	&	---	&	---	\\
L2	&	10	&	0	&	27	&	27	&	16	&	16	&	---	&	---	\\
L3	&	12	&	3	&	32	&	38	&	20	&	46	&	---	&	---	\\
L4	&	15	&	1	&	27	&	27	&	20	&	28	&	---	&	---	\\
\enddata
\tablecomments{The age range is calculated from the Wielen (1977) age-velocity relation for the disk which uses a value of $\alpha$ of (1/3).}

\end{deluxetable}

\pagebreak

\begin{deluxetable}{llccccccc}
\tabletypesize{\scriptsize}
\label{tab:tab1}
\tablecaption{Full Astrometric Sample}
\tablewidth{0pt}
\tablehead{
\colhead{SpT} &
\colhead{N} &
\colhead{N } &
\colhead{median $V_{tan}$} &
\colhead{median $V_{tan}$} &
\colhead{$\sigma_{tan}$} &
\colhead{$\sigma_{tan}$} &
\colhead{Age} &
\colhead{Age} \\
& &
\colhead{High $V_{tan}$}&
&
\colhead{with high $V_{tan}$}&
&
\colhead{with high $V_{tan}$}&
&
\colhead{with high $V_{tan}$}\\
 & & & \colhead{(km~s$^{-1}$)} &
\colhead{(km~s$^{-1}$)} &
\colhead{(km~s$^{-1}$)} &
\colhead{(km~s$^{-1}$)} &
\colhead{(Gyr)} &
\colhead{(Gyr)} \\
\colhead{(1)} &
\colhead{(2)} &
\colhead{(3)} &
\colhead{(4)} &
\colhead{(5)} &
\colhead{(6)} &
\colhead{(7)} &
\colhead{(8)} &
\colhead{(9)} \\
}
\startdata

M7	&	88	&	0	&	27	&	27	&	19	&	19	&	---	&	---	\\
M8	&	114	&	1	&	27	&	27	&	21	&	23	&	---	&	---	\\
M9	&	71	&	1	&	23	&	23	&	19	&	21	&	---	&	---	\\
L0	&	93	&	1	&	19	&	19	&	16	&	21	&	---	&	---	\\
L1	&	83	&	2	&	32	&	33	&	23	&	27	&	---	&	---	\\
L2	&	58	&	0	&	26	&	26	&	18	&	18	&	---	&	---	\\
L3	&	64	&	3	&	30	&	32	&	18	&	27	&	---	&	---	\\
L4	&	50	&	1	&	25	&	27	&	20	&	23	&	---	&	---	\\
\enddata
\tablecomments{The age range is calculated from the Wielen (1977) age-velocity relation for the disk which uses a value of $\alpha$ of (1/3).}

\end{deluxetable}

\pagebreak

\begin{deluxetable}{rrcccccr}
\label{tab:tab6}
\tabletypesize{\scriptsize}
\tablecaption{ Average Kinematics and Ages for the Subgroups}
\tablewidth{0pt}
\tablehead{
\colhead{SpT} &
\colhead{N} &
\colhead{Median $V_{tan}$} &
\colhead{ $\sigma_{tan}$} &
\colhead{Age Range} \\
 & &
\colhead{km~s$^{-1}$} &
\colhead{km~s$^{-1}$} &
\colhead{Gyr} \\
\colhead{(1)} &
\colhead{(2)} &
\colhead{(3)} &
\colhead{(4)} &
\colhead{(5)} \\
}
\startdata
 M7-T9/BLUE	&	16	&	53	&	47	&	$37.9^{+12.6}_{-10.3}$	\\
M7-T9/RED	&	29	&	26	&	16	&	$1.2^{+0.5}_{-0.4}$	\\
			\hline
 \hline						
M7-L9/BLUE	&	13	&	56	&	50	&	$46.0^{+15.2}_{-12.4}$	\\
M7-L9/RED	&	24	&	26	&	15	&	$1.0^{+0.4}_{-0.3}$	\\
			\hline
 \hline						
 UBLs	&	10	&	99	&	47	&	$37.9^{+12.6}_{-10.3}$	\\
Low Gravity	&	26	&	18	&	15	&	$1.0^{+0.4}_{-0.3}$	\\      \hline
      \hline
	\enddata
\tablecomments{The age range is calculated from the Wielen (1977) age-velocity relation for the disk which uses a value of $\alpha$ of (1/3).}
\end{deluxetable}

\begin{deluxetable}{l l l l l l l l l l l l l l l l}
\label{tab:tab7}
\tabletypesize{\tiny}
\tablecaption{ Details on Red Photometric Outliers}
\tablewidth{0pt}
\small
\tablehead{
\colhead{Source Name} &
\colhead{2MASS J} &
\colhead{2MASS $K_{s}$} &
\colhead{$\mu_{\alpha}cos(\delta)$ } &
\colhead{$ \mu_{\delta}$}  &
\colhead{$\mu$ Ref.} &
\colhead{SpT} &
\colhead{SpT}&
\colhead{$V_{tan}$}&
\colhead{Note\tablenotemark{r}}  \\

 & 
\colhead{(mag)} &
\colhead{ (mag)} &
\colhead{($\arcsec$/yr)} &
\colhead{(\arcsec/yr)} &
&
\colhead{(opt)} &
\colhead{(IR)} &
\colhead{(km~s$^{-1}$)} \\

\colhead{(1)} &
\colhead{(2)} &
\colhead{(3)} &
\colhead{(4)} &
\colhead{(5)} &
\colhead{(6)} &
\colhead{(7)} &
\colhead{(8)} &
\colhead{(9)} &
\colhead{(11)} \\

}
\startdata 
2MASS J00374306-5846229	&	15.37	$\pm$	0.05		&	13.59	$\pm$	0.05		&	0.049	$\pm$	0.010	&	-0.051	$\pm$	0.020	&	19	&	L0	&	---	&	18	$\pm$	5	&	LG	\\
SDSSp J010752.33+004156.1	&	15.82	$\pm$	0.06		&	13.71	$\pm$	0.04		&	0.628	$\pm$	0.007	&	0.091	$\pm$	0.004	&	44	&	L8	&	L5.5	&	46.9	$\pm$	3.3	&	---	\\
2MASS J01244599-5745379	&	16.31	$\pm$	0.11		&	14.32	$\pm$	0.09		&	-0.003	$\pm$	0.010	&	0.018	$\pm$	0.019	&	19	&	L0	&	---	&	7	$\pm$	7	&	LG	\\
2MASS J01415823-4633574	&	14.83	$\pm$	0.04		&	13.10	$\pm$	0.03		&	0.104	$\pm$	0.017	&	-0.026	$\pm$	0.024	&	19	&	L0	&	L0	&	21	$\pm$	4	&	LG	\\
2MASS J01490895+2956131	&	13.45	$\pm$	0.02		&	11.98	$\pm$	0.02		&	0.1757	$\pm$	0.0008	&	-0.4021	$\pm$	0.0007	&	15	&	M9.5	&	---	&	46.8	$\pm$	0.7	&	---	\\
2MASSI J0243137-245329	&	15.42	$\pm$	0.06\tablenotemark{c}	&	15.22	$\pm$	0.06\tablenotemark{c}	&	-0.288	$\pm$	0.004	&	-0.208	$\pm$	0.003	&	44	&	---	&	T6	&	18.0	$\pm$	0.7	&	---	\\
2MASS J03231002-4631237	&	15.39	$\pm$	0.07		&	13.70	$\pm$	0.05		&	0.060	$\pm$	0.013	&	-0.010	$\pm$	0.019	&	19	&	L0	&	---	&	16	$\pm$	4	&	LG	\\
2MASS J03264225-2102057	&	16.13	$\pm$	0.09		&	13.92	$\pm$	0.07		&	0.108	$\pm$	0.014	&	-0.146	$\pm$	0.015	&	19	&	L4	&	---	&	35	$\pm$	5	&	---	\\
\enddata
\tablecomments{See Table 3 for references and notes referred to in this table.}
\end{deluxetable}

\begin{deluxetable}{l l l l l l l l l l l l l l l l}
\label{tab:tab7}
\tabletypesize{\tiny}
\tablecaption{ Details on Blue Photometric Outliers}
\tablewidth{0pt}
\small
\tablehead{
\colhead{Source Name} &
\colhead{2MASS J} &
\colhead{2MASS $K_{s}$} &
\colhead{$\mu_{\alpha}cos(\delta)$ } &
\colhead{$ \mu_{\delta}$}  &
\colhead{$\mu$ Ref.} &
\colhead{SpT} &
\colhead{SpT}&
\colhead{$V_{tan}$}&
\colhead{Note\tablenotemark{r}}  \\

 & 
\colhead{(mag)} &
\colhead{ (mag)} &
\colhead{($\arcsec$/yr)} &
\colhead{(\arcsec/yr)} &
&
\colhead{(opt)} &
\colhead{(IR)} &
\colhead{(km~s$^{-1}$)} \\

\colhead{(1)} &
\colhead{(2)} &
\colhead{(3)} &
\colhead{(4)} &
\colhead{(5)} &
\colhead{(6)} &
\colhead{(7)} &
\colhead{(8)} &
\colhead{(9)} &
\colhead{(11)} \\

}
\startdata 
HD 3651B	&	16.16	$\pm$	0.03		&	16.87	$\pm$	0.05		&	-0.4611	$\pm$	0.0007	&	-0.3709	$\pm$	0.0007	&	34	&	---	&	T7.5	&	31.2	$\pm$	0.3	&	VLMC	\\
SSSPM J0134-6315	&	14.51	$\pm$	0.04		&	13.70	$\pm$	0.04		&	0.077	$\pm$	0.008	&	-0.081	$\pm$	0.009	&	30	&	---	&	L0	&	19	$\pm$	2	&	---	\\
2MASS J02530084+1652532	&	8.39	$\pm$	0.03		&	7.59	$\pm$	0.05		&	3.404	$\pm$	0.005	&	-3.807	$\pm$	0.005	&	23	&	M7	&	---	&	92.9	$\pm$	1.0	&	---	\\
SDSS J090900.73+652527.2	&	16.00	$\pm$	0.06\tablenotemark{a}	&	15.16	$\pm$	0.06\tablenotemark{a}	&	-0.217	$\pm$	0.003	&	-0.138	$\pm$	0.008	&	19	&	---	&	T1	&	28	$\pm$	1	&	---	\\
2MASS J09211410-2104446	&	12.78	$\pm$	0.02		&	11.69	$\pm$	0.02		&	0.244	$\pm$	0.016	&	-0.908	$\pm$	0.017	&	19	&	L2	&	---	&	56	$\pm$	4	&	UBL	\\
SDSS J093109.56+032732.5	&	16.75	$\pm$	0.10\tablenotemark{c}	&	15.65	$\pm$	0.10\tablenotemark{c}	&	-0.612	$\pm$	0.018	&	-0.131	$\pm$	0.018	&	19	&	---	&	L7.5	&	108	$\pm$	23	&	UBL	\\
2MASSI J0937347+293142	&	14.58	$\pm$	0.06\tablenotemark{c}	&	15.51	$\pm$	0.12\tablenotemark{c}	&	0.973	$\pm$	0.005	&	-1.298	$\pm$	0.006	&	44	&	d/sdT6	&	T6	&	47.2	$\pm$	1.1	&	---	\\
SDSS J103321.92+400549.5	&	16.88	$\pm$	0.06\tablenotemark{a}	&	15.63	$\pm$	0.10\tablenotemark{a}	&	0.154	$\pm$	0.013	&	-0.188	$\pm$	0.018	&	19	&	---	&	L6	&	53	$\pm$	10	&	UBL	\\
\enddata
\tablecomments{See Table 3 for references and notes referred to in this table.}
\end{deluxetable}

\begin{deluxetable}{r r r r c c c r r r r r | c c}
\label{tab:tab7}
\tabletypesize{\scriptsize}
\rotate
\tablecaption{ High $V_{tan}$ Objects}
\tablewidth{0pt}
\tablehead{
\colhead{Discovery Name} &
\colhead{J-$K_{s}$} &
\colhead{2MASS J} &
\colhead{2MASS $K_{s}$} &
\colhead{$\mu_{\alpha}cos(\delta)$ } &
\colhead{$ \mu_{\delta}$}  &
\colhead{SpT} &
\colhead{SpT}&
\colhead{Distance}&
\colhead{$V_{tan}$}&
\colhead{Note\tablenotemark{r}}  \\

 & &
\colhead{(mag)} &
\colhead{ (mag)} &
\colhead{($\arcsec$/yr)} &
\colhead{(\arcsec/yr)} &
\colhead{(opt)} &
\colhead{(IR)} &
\colhead{(pc)} &
\colhead{(km~s$^{-1}$)} \\

\colhead{(1)} &
\colhead{(2)} &
\colhead{(3)} &
\colhead{(4)} &
\colhead{(5)} &
\colhead{(6)} &
\colhead{(7)} &
\colhead{(8)} &
\colhead{(9)} &
\colhead{(10)} &
\colhead{(11)} \\

}
\startdata
DENIS-P J1253108-570924	&	1.40	&	13.45	$\pm$	0.02		&	12.05	$\pm$	0.02		&	-1.575	$\pm$	0.005	&	-0.434	$\pm$	0.014	&	L0.5	&	---	&	21	$\pm$	3		&	162	$\pm$	20	&	---	\\
2MASSIJ1721039+334415	&	1.14	&	13.63	$\pm$	0.02		&	12.49	$\pm$	0.02		&	-1.854	$\pm$	0.017	&	0.602	$\pm$	0.017	&	L3	&	---	&	16	$\pm$	1		&	144	$\pm$	13	&	UBL	\\
2MASSJ11145133-2618235	&	-0.25	&	15.86	$\pm$	0.08		&	$<$ 16.11				&	-3.03	$\pm$	0.04	&	-0.36	$\pm$	0.04	&	---	&	T7.5	&	10	$\pm$	2		&	140	$\pm$	22	&	---	\\
2MASSWJ1411175+393636	&	1.40	&	14.64	$\pm$	0.03		&	13.24	$\pm$	0.04		&	-0.911	$\pm$	0.015	&	0.137	$\pm$	0.016	&	L1.5	&	---	&	32	$\pm$	2		&	138	$\pm$	10	&	---	\\
2MASS J182611.31+301420.1	&	0.85	&	11.66	$\pm$	0.02		&	10.81	$\pm$	0.02		&	-2.280	$\pm$	0.010	&	-0.684	$\pm$	0.010	&	M8.5	&	---	&	12	$\pm$	1		&	132	$\pm$	9	&	---	\\
2MASSJ21501592-7520367	&	1.38	&	14.06	$\pm$	0.03		&	12.67	$\pm$	0.03		&	0.980	$\pm$	0.048	&	-0.281	$\pm$	0.014	&	L1	&	---	&	26	$\pm$	3		&	125	$\pm$	18	&	---	\\
2MASSIJ0251148-035245	&	1.40	&	13.06	$\pm$	0.03		&	11.66	$\pm$	0.02		&	1.128	$\pm$	0.013	&	-1.826	$\pm$	0.020	&	L3	&	L1	&	12	$\pm$	1		&	122	$\pm$	11	&	---	\\
SDSS J133148.92-011651.4	&	1.35	&	15.48	$\pm$	0.06\tablenotemark{c}	&	14.12	$\pm$	0.06\tablenotemark{c}	&	-0.407	$\pm$	0.019	&	-1.030	$\pm$	0.014	&	L6	&	L8	&	23	$\pm$	2		&	119	$\pm$	11	&	UBL	\\
\enddata
\tablecomments{See Table 3 for references and notes referred to in this table.}
\end{deluxetable}

\begin{deluxetable}{ccccccccccccccc}
\label{tab:tab6}
\tabletypesize{\scriptsize}
\tablecaption{ Median Kinematics and Ages for the 20 pc sample of nearby Stars}
\rotate

\tablewidth{0pt}
\tablehead{
\colhead{SpT} &
\colhead{N\tablenotemark{a}} &
\colhead{ U} &
\colhead{ $\sigma_{U}$} &
\colhead{ V} &
\colhead{ $\sigma_{V}$} &
\colhead{ W} &
\colhead{ $\sigma_{W}$} &
\colhead{ $V_{tan}$}&
\colhead{ $\sigma_{tan}$}&
\colhead{ Age from $V_{tan}$} &
\colhead{N\tablenotemark{b} }&
\colhead{ $V_{tot}$}&
\colhead{ $\sigma_{tot}$}&
\colhead{ Age from $V_{tot}$} \\
 & &
\colhead{(km/s)} &
\colhead{(km/s)} &
\colhead{(km/s)} &
\colhead{(km/s)} &
\colhead{(km/s)} &
\colhead{(km/s)} &
\colhead{(km/s)} &
\colhead{(km/s)} &
\colhead{(Gyr)}  &
&
\colhead{(km/s)} &

\colhead{(km/s)} &
\colhead{(Gyr)}  \\
\colhead{(1)} &
\colhead{(2)} &
\colhead{(3)} &
\colhead{(4)} &
\colhead{(5)} &
\colhead{(6)} &
\colhead{(7)} &
\colhead{(8)} &
\colhead{(9)} &
\colhead{(10)} &
\colhead{(11)} &
\colhead{(12)} &
\colhead{(13)} &
\colhead{(14)} &
\colhead{(15)} \\

}
\startdata
F	&	139	&	-6	&	28	&	-8	&	18	&	-6	&	16	&	25	&	17	&	$	1.7	^{+	0.6	}_{-	0.5	}$	&	139	&	34	&	19	&	$	1.2	^{+	0.5	}_{-	0.4	}$	\\
G	&	221	&	-16	&	38	&	-15	&	29	&	-5	&	21	&	32	&	27	&	$	7.3	^{+	2.5	}_{-	2.0	}$	&	221	&	40	&	30	&	$	5.4	^{+	1.9	}_{-	1.5	}$	\\
K	&	308	&	-13	&	42	&	-20	&	30	&	-8	&	18	&	35	&	28	&	$	7.8	^{+	2.6	}_{-	2.2	}$	&	308	&	46	&	30	&	$	5.2	^{+	1.8	}_{-	1.5	}$	\\
M0-M6	&	60	&	-8	&	45	&	-14	&	22	&	-7	&	21	&	32	&	30	&	$	9.5	^{+	3.2	}_{-	2.6	}$	&	60	&	38	&	31	&	$	6.0	^{+	2.0	}_{-	1.7	}$	\\
M7-M9	&	93	&	-6	&	26	&	-14	&	20	&	-7	&	20	&	29	&	24	&	$	5.0	^{+	1.7	}_{-	1.4	}$	&	81	&	28	&	21	&	$	1.7	^{+	0.6	}_{-	0.5	}$	\\
L0-L9	&	114	&	-8	&	28	&	-15	&	23	&	-5	&	15	&	27	&	26	&	$	6.6	^{+	2.2	}_{-	1.8	}$	&	168	&	25	&	24	&	$	2.6	^{+	0.9	}_{-	0.7	}$	\\
T0-T8	&	70	&	-8	&	33	&	-12	&	21	&	-8	&	16	&	31	&	24	&	$	4.6	^{+	1.6	}_{-	1.3	}$	&	35	&	23	&	23	&	$	2.2	^{+	0.8	}_{-	0.7	}$	\\
\enddata
	\tablenotetext{a}{The number of objects used to calculate median $V_{tan}$ values.}
	\tablenotetext{b}{The number of objects used to calculate median $V_{tot}$ values and thus used in the U,V,W analysis.}
	
\tablecomments{Kinematic data for F,G,K, and early M stars gathered from the \citet{2003A&A...398..141S}, \citet{2004yCat.3239....0K} and \citet{2004A&A...418..989N} catalogues.  We restricted to distances $<$ 20 pc and proper motions $>$ 20 mas/yr  for comparison with our 20 pc ultracool dwarf sample.}
\end{deluxetable}

\pagebreak
\pagebreak

\end{document}